\def\year{2022}\relax
\documentclass[letterpaper]{article} 
\usepackage{aaai22}  
\usepackage{times}  
\usepackage{helvet}  
\usepackage{courier}  
\usepackage[hyphens]{url}  
\usepackage{graphicx} 
\usepackage{natbib}  
\usepackage{caption} 
\DeclareCaptionStyle{ruled}{labelfont=normalfont,labelsep=colon,strut=off} 
\usepackage{algorithm}
\usepackage{algorithmic}
\usepackage{subcaption}
\usepackage{bm}
\usepackage{soul}
\usepackage{xcolor}

\usepackage[dvipsnames]{xcolor}
\definecolor{ForestGreen}{rgb}{0.13, 0.55, 0.13}
\definecolor{VibrantBlue}{rgb}{0.0, 0.2, 1.0} 

\newcommand{\new}[1]{\textcolor{black}{#1}}
\definecolor{DarkMagenta}{rgb}{0.55, 0, 0.55}

\newcommand{\fm}[1]{\textcolor{black}{#1}}
\setstcolor{red}

\usepackage{newfloat}
\usepackage{listings}
\floatstyle{ruled}
\newfloat{listing}{tb}{lst}{}
\floatname{listing}{Listing}

\usepackage[utf8]{inputenc}
\usepackage{graphicx}
\usepackage{multirow} 
\usepackage{booktabs}
\usepackage{newunicodechar}
\newunicodechar{↳}{\rotatebox[origin=c]{180}{$\Lsh$}}
\usepackage{array}
\usepackage{amsmath}
\usepackage{xcolor}
\newcommand{\answerYes}[1]{\textcolor{blue}{#1}} 
 
\newcommand{\answerNA}[1]{\textcolor{gray}{#1}} 
 
\title{Political Advertising on Facebook During the 2022 Australian Federal Election: \\A Social Identity Perspective}

\author{
Stefano Civelli, Pietro Bernardelle, Frank Mols, Gianluca Demartini\\
}
\affiliations {
The University of Queensland, Brisbane, Australia\\
\{s.civelli,p.bernardelle,f.mols,g.demartini\}@uq.edu.au
}

\usepackage{bibentry}

\begin{document}
\maketitle
\begin{abstract}
The spread of targeted advertising on social media platforms has revolutionized political marketing strategies. 
Monitoring these digital campaigns is essential for maintaining transparency and accountability in democratic processes.
Leveraging Meta's Ad Library, we analyze political advertising on Facebook and Instagram during the 2022 Australian federal election campaign. 
We investigate temporal, demographic, and geographical patterns in the advertising strategies of major Australian political actors to establish an empirical evidence base, and interpret these findings through the lens of Social Identity Theory (SIT). Our findings not only reveal significant disparities in spending and reach among parties, \fm{but also in persuasion strategies being deployed in targeted online campaigns: appeals to material self-interests, eliciting subconscious automatic pilot responses, and attempts to harness social identities.}
We observe a marked increase in advertising activity as the election approached, peaking just before the mandated media blackout period. Demographic analysis shows distinct targeting strategies, with parties focusing more on younger demographics and exhibiting gender-based differences in ad impressions. Regional distribution of ads largely mirrored population densities, with some parties employing more targeted approaches in specific states. Moreover, we found that parties emphasized different themes aligned with their ideologies—major parties focused on party names and opponents, while smaller parties emphasized issue-specific messages. 
\new{Drawing on SIT, we interpret these findings within Australia's compulsory voting context, suggesting that parties employed distinct persuasion strategies. With turnout guaranteed, major parties focused on reinforcing partisan identities to prevent voter defection, while smaller parties cultivated issue-based identities to capture the support of disaffected voters who are obligated to participate.}
\end{abstract}

\section{Introduction}
Research on social media's impact on politics has evolved over time. Early studies focused on Twitter's role in reflecting public opinion and predicting election outcomes \cite{o2010tweets,tumasjan2011election}.
However, subsequent studies identified several shortcomings in these initial approaches, pointing to issues with methodology and representative sampling across demographic groups \cite{gayo2012wanted,gayo2012no,metaxas2011not}.
As the research landscape matured, researchers began examining the dual impact of social media on political campaigns—its ability to enable precise audience targeting and efficient campaign spending \cite{fowler2021political} through precise audience segmentation and dynamic campaign adjustments \cite{boerman2017online,kreiss2016prototype}, alongside growing concerns about disinformation and manipulation of public opinion \cite{entous2017russian}.
Facebook's emergence as a critical platform for political advertising marked a significant shift in this research trajectory. In 2018, Meta (formerly Facebook) launched its Ad Library in response to transparency concerns, providing unprecedented access to political advertisement data. This development has sparked new research directions examining Facebook's role in political campaigning across various contexts \cite{calvo2021global,capozzi2021clandestino,mejova2020covid,pierri2022itaelec}. 

\begin{figure}[t]
    \centering
    \includegraphics[width=1\linewidth]{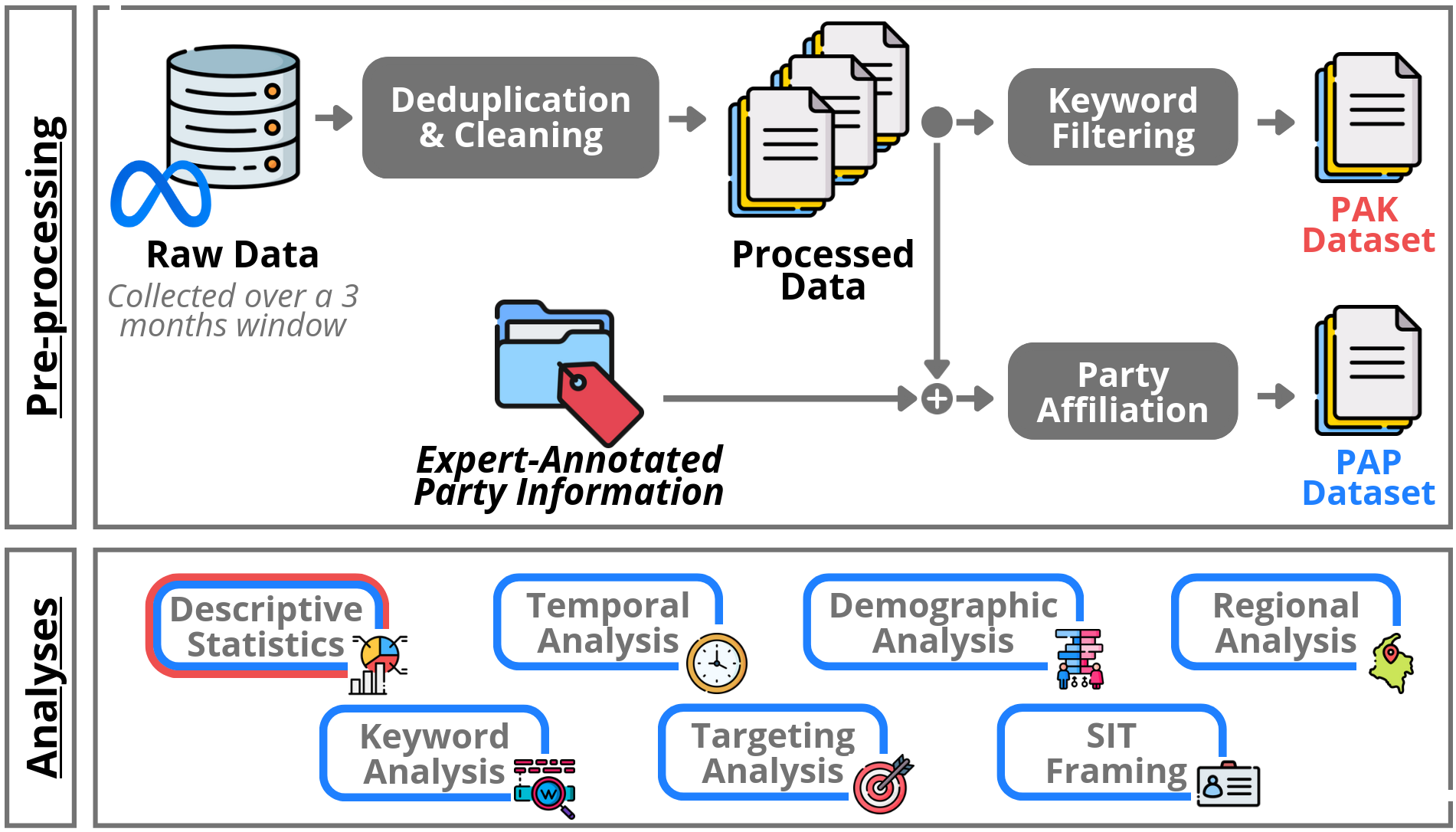}
    \caption{Pipeline for collecting, cleaning, and annotating Meta Ad Library data, followed by analyses.}
    \label{fig:overview}
\end{figure}

\new{However, most research focuses on voluntary voting systems, where mobilizing a party’s base to turn out is a central objective. By contrast, Australia’s system of compulsory voting ensures consistently high participation\footnote{https://results.aec.gov.au/27966/Website/HouseTurnoutByState-27966.htm}, shifting the strategic focus from mobilization—or “Get Out The Vote” (GOTV) efforts—toward persuasion \cite{jackman2001compulsory, Birch2009compulsory}. Rather than expending resources to secure turnout, campaigns concentrate on convincing undecided voters and reinforcing party loyalty, incentivizing broad-based appeals over the micro-targeted mobilization common in voluntary systems \cite{hillygus2008persuadable}. 
This institutional feature makes Australia a distinctive setting to study digital advertising tactics and often encourages parties to moderate their messaging as they compete for the median voter \cite{jackman2001compulsory,bruns2018social,skoric2012tweets, oprea2024moving}.}

Our study addresses this empirical knowledge gap by analyzing political advertisements on Meta platforms (Facebook and Instagram) during the 2022 Australian federal election, and \fm{by evaluating the findings using three well-established political psychology perspectives on persuasion, namely (a) appealing to rational self-interest, (b) eliciting automatic pilot behavior, and (c) social identity mobilization.}
By curating a comprehensive dataset, annotating party affiliations, and examining spending, targeting, and content strategies (Figure~\ref{fig:overview}), we provide insights into digital campaign dynamics in a unique electoral system, building on prior work while addressing platform-specific and context-specific nuances \cite{jungherr2016twitter,skoric2012tweets}. 
Accordingly, our research seeks to answer the following questions:

\begin{itemize}
    \item \textbf{RQ1.} \textit{What was the scale and reach of political advertisements on Meta platforms during the election campaign?} 
    \item \textbf{RQ2.} \textit{How did the advertising strategies of major Australian political parties and groups differ in terms of temporal distribution of ads, spending patterns, targeting and the use of themes throughout the campaign period?}
    \item \textbf{RQ3.} \textit{How can the observed patterns in advertising content and targeting strategies be interpreted through theoretical frameworks of voter motivation?}
\end{itemize}

\paragraph{Contributions.}
Our work contributes to the understanding of digital political advertising in the following ways: \textbf{(1)} we curated a dataset of political advertisements from the election period; \textbf{(2)} we had a political science expert annotate a list of parties to categorize ads by political affiliation; \textbf{(3)} provide a general analysis of the campaign, comparable to studies conducted in other countries; and \textbf{(4)} offer insights into major parties' online campaign strategies, examining both quantitative aspects (spending, timing, targeting) and the theoretical frameworks that may explain their persuasion approaches.

\paragraph{Findings.}
The Australian Labor Party (ALP)\footnote{We use the terms ``ALP'' and ``Labor'' interchangeably.} and the Liberal Coalition\footnote{The Coalition comprises the Liberal Party, National Party, Liberal National Party QLD, and Country Liberal NT.} dominated the digital advertising landscape, collectively accounting for over 70\% of total ad impressions and spending. The ALP invested A\$5M (40.2\% of total political ad spending) across 13,047 ads, while the Liberal Coalition spent A\$3.9M (31.3\% of total political ad spending) across 9,364 ads (\textbf{RQ1}). These two major parties secured 68.27\% of the electors' first preference votes on election day.

We observed distinct patterns in advertising strategies throughout the election timeline (\textbf{RQ2}). After the April 10th election call, ad activity surged. Labor consistently outspent the Liberal Coalition, though both intensified efforts toward the campaign's end. The Greens and UAP followed different timelines, with UAP ramping up late. Labor targeted more women across age groups; the Greens skewed younger; UAP focused on male audiences. 
Geographically, major parties followed population centers, while smaller parties used more targeted regional strategies. Major parties used ads to promote themselves and attack opponents, while smaller parties emphasized specific issues and policy themes.

When viewed through Social Identity Theory (SIT) (\textbf{RQ3}), these findings suggest strategic differentiation in identity appeals. Established major parties worked to strengthen and mobilize existing partisan identifications, whereas smaller political entities sought to foster and nurture identity connections based on specific policy concerns and issue salience. This theoretical interpretation helps explain the observed variations in targeting strategies, with parties tailoring their approaches to activate different identity-based motivations among potential voters.

\section{Background and Related Work}

\subsection{Theoretical Frameworks of Voter Motivation}
\label{ss:politicalTheory}

Political scientists have long explained voting behavior through (egoistic or sociotropic) rational self-interest, a view rooted in early-20th-century economic thought. This began to shift in the 1950s when researchers like \citet{simon1957models} introduced the concept of ‘bounded rationality’, showing that individuals often rely on heuristics under cognitive strain. This opened the door for insights from cognitive psychology—such as ‘groupthink’, nudging, and behavioral public policy—though these have yet to significantly influence psephology (election studies). This is surprising, given that many campaign messages clearly aim to trigger automatic emotional responses rather than deliberative reasoning, as seen in the Liberal party’s Frankencredit\footnote{https://independentaustralia.net/politics/politics-display/franking-credits-the-coalitions-scare-campaign,12365} and Labor’s Mediscare\footnote{https://theconversation.com/why-scare-campaigns-like-mediscare-work-even-if-voters-hate-them-62279} campaigns.
One psychological theory that has been incorporated in election studies, albeit belatedly, is SIT \cite{tajfel1978achievement,tajfel1979integrative}, which emphasizes partisan identity as a key driver of vote choice. This aligns with \citet{campbell1960american}, who found that most voters are firmly loyal to their chosen party—what would be described in Australia as ‘rusted-on’ supporters. Recent interest in SIT has grown alongside rising affective polarization \cite{garrett2020moral} and negative partisanship \cite{abramowitz2018negative,bankert2020origins}, with strategists increasingly using ‘us-them’ narratives to build partisan identities \cite{huddy2017political,west2022partisanship}. However, this literature is still largely absent from digital campaigning research.\footnote{Two major models in this area are ‘Expressive Partisanship’—viewing partisanship as emotional group attachment \cite{green2004partisan,huddy2017political}—and ‘Instrumental Partisanship’, which sees it as a rational attachment based on party performance \cite{fiorina1981retrospective,orr2020policy}. In practice, these processes likely interact, making it more useful to consider how different persuasion strategies draw on self-interest, cognition, or identity.}

\new{This identity-based perspective is particularly salient in Australia’s compulsory voting system. Because turnout is high, parties focus less on mobilization and more on persuasion \cite{jackman2001compulsory, Birch2009compulsory}. This environment encourages appeals aimed at the median voter, pushing parties to focus on broadly popular messages while avoiding highly salient or polarizing issues \cite{oprea2024moving}. Over time, such strategies can reinforce partisan loyalties. Cross-national research suggests that compulsory voting further strengthens habitual partisan identities \cite{singh2013compulsory}.
As partisan identity becomes more central to vote choice \cite{huddy2015partisan}, parties can either positively reinforce the in-group or rely on negative partisanship—defining themselves against an out-group—to strengthen these attachments \cite{abramowitz2018negative,bankert2020origins}}.

In sum, there is significant opportunity for interdisciplinary research into why particular campaign messages resonate. We suggest that insights from three psychological models—(a) material self-interest, (b) flawed cognition, and (c) partisan social identity—could help explain why particular tactics hit the mark in digital political campaigns.

\subsection{Social Media Political Advertising}
Digital channels have fundamentally altered modern political messaging \cite{fowler2021political}.
These new platforms offer distinct advantages for political campaigns, including unprecedented audience targeting precision and cost-effectiveness compared to traditional media \cite{boerman2017online,kreiss2016prototype}. Building on this foundation, \citet{dommett2019political} explored the political economy of Facebook advertising, documenting its growing prominence in election spending and strategic campaign planning.

The introduction of the Meta Ad Library in 2018 has enabled researchers to examine Facebook's role in political advertising across diverse contexts \cite{calvo2021global,capozzi2021clandestino,leerssen2021news,mejova2020covid,pierri2022itaelec}. 
\new{Recent studies have also introduced weakly and minimally supervised machine learning methods to classify ad stances, themes, and moral foundations in political and issue-based campaigns, such as those on elections, climate change, and COVID-19 vaccines, revealing targeting patterns and narrative strategies \cite{islam2023weakly,islam2023climate,islam2022covid}.}

These studies have uncovered both new campaign opportunities and significant transparency concerns related to platform-based political messaging \cite{entous2017russian}\new{, including imprecise moderation with high false-positive and false-negative rates that allow undeclared coordinated campaigns to evade detection \cite{bouchaud2024beyond}.}
This recent work extends earlier research that examined Twitter's potential for reflecting public opinion and predicting electoral outcomes \cite{o2010tweets,tumasjan2011election}, though these approaches faced substantial methodological challenges and critiques \cite{gayo2012wanted,gayo2012no,metaxas2011not}.

The evolution of digital political advertising has increasingly centered on the strategic deployment of micro-targeting techniques, which leverage the granular demographic and behavioral data available on social media platforms. This shift represents a significant departure from traditional broadcast approaches to political messaging. \citet{ridout2021influence} investigated how political campaigns deploy ads on Facebook, focusing on the influence of campaign goals and timing. Their work provides valuable insights into the strategic considerations behind digital ad placement.
\citet{brodnax2022home} studied the evolution of digital advertising strategies during the 2020 US Presidential Primary, highlighting the shift from broad to more targeted approaches as campaigns progressed. \new{Recent analyses have further revealed systematic discrepancies between targeted and actual ad deliveries, with algorithmic biases favoring certain parties and skewing audiences by age and gender \cite{bar2024systematic}.}

While our study focuses on a different political context, these established methodological approaches inform our analysis of partisan advertising strategies in Australia.

\subsection{Australian Context}
While numerous studies have examined political advertising on social media in various countries, research specific to the Australian context is relatively limited. \citet{bruns2018social} provided an analysis of Twitter usage during the 2013 and 2016 Australian federal elections, finding significant differences in how major parties utilized the platform. They noted that Labor candidates were consistently more active on Twitter than their Liberal Coalition counterparts across both elections, a trend that aligns with our findings.
More recently, \citet{DeckerTopic} explored topic diversity in social media campaigning during the 2022 Australian federal election. Their study found that while there were some differences between organic and paid content, both types predominantly focused on core political topics aligned with party ideologies. They also noted the emergence of government critique as a distinct topic in both organic and paid content, signaling the use of negative campaigning.

\new{While \citet{arya2022political} and \citet{meguellati2025} also examined political advertising during the 2022 Australian Federal election, our work extends theirs in several ways. We release a manually curated party affiliation list to support reproducibility, analyze an extra month of data with more advanced analysis methods (including keyword filtering and targeting) and frame the findings through the lens of SIT.}

\subsection{The 2022 Australian Federal Election}
The 2022 Australian federal election was called on April 10th, with election day set for May 21st. Approximately 17.2 million eligible voters were registered to participate. The election resulted in a change of government, with the Australian Labor Party (ALP) securing 77 seats in the 151-seat House of Representatives, allowing them to form a majority government. The Liberal-National Coalition, which had been in power since 2013, won 58 seats. Minor parties and independents claimed the remaining 16 seats, 12 more as compared to the previous election, reflecting a growing trend towards diversification in Australian politics. The ALP received 32.6\% of the primary votes, while the Coalition garnered 35.7\%. However, after preference distribution under Australia's two-party-preferred system, these figures adjusted to 52.1\% and 47.9\% respectively. The Greens achieved their best-ever result with 12.2\% of the primary votes, followed by independent candidates with 5.3\%. This election was notable for the significant swing away from the two major parties, with their combined primary vote falling below 70\% for the first time in Australian history.

\section{Data Preparation}
\label{s:data_curation}

\subsection{Data Collection}

We utilized the Meta Ad Library API\footnote{\url{https://www.facebook.com/ads/library/}} to collect online advertisements published by parties, political candidates, and other entities related to the 2022 Australian federal election. Our data collection process captured actively running advertisement campaigns at six-hour intervals from March 1\textsuperscript{st}, 2022, to June 1\textsuperscript{st}, 2022. This resulted in a dataset of 56,224 ads from 1,549 sponsors.

Each ad includes attributes such as the creation time, target demographic distributions, ad text, page name, sponsor, ad URL, and metrics of ad impression and spending. For these last two, the Meta Ad Library API provides upper and lower bounds. In our analysis, we calculated ad spend and impressions as the mean of these bounds. 

We determined impressions by age, region, and gender by multiplying the mean impressions for each ad by the proportion of ad reach attributed to each category. Ads lacking distribution data for age, gender, or region were excluded from calculations for those specific impression categories. The 13–17 age group and users with an "unknown" gender were excluded from the demographic analysis due to their minimal presence, accounting for only 0.1\% and 1.1\% of ad impressions, respectively.

It is important to note that our initial dataset, PA (Political Advertisements), comprised only ads explicitly flagged as political, as required by Australian law\footnote{In Australia the posting entity is legally required to flag political ads with a ``paid by'' disclaimer.}. This approach may introduce potential biases, as some political ads might not have been flagged and thus excluded, while others could have been mis-flagged as political when they were not, a phenomenon studied by \citet{complexity_political_ads_detection}. 
\new{Reinforcing this concern, \citet{bouchaud2024beyond} studied Meta's EU ad moderation and estimate that 92.3\% of undeclared political ads (those not flagged as political by the publisher) are not detected by Meta's systems, highlighting a critical loophole in its reliance on voluntary self-declaration by advertisers}

\begin{figure}[t]
    \centering
    \includegraphics[width=1.41\linewidth, angle=270]{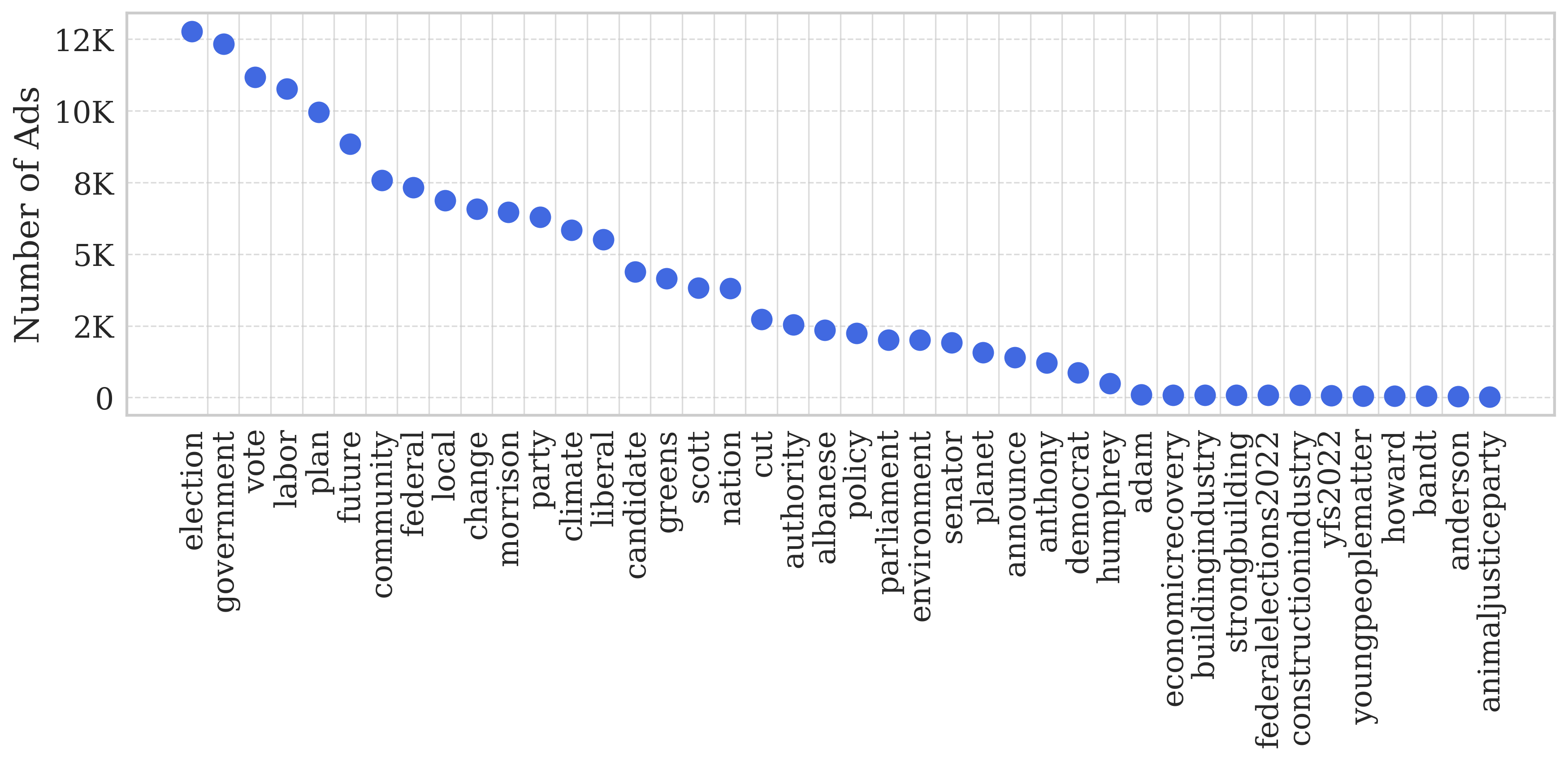}
    \caption{Frequency distribution of selected keywords used in political advertisements during the considered period. Keywords are ranked by the number of ads they appear in.}
    \label{fig:keywords}
\end{figure}

\subsection{Data Preprocessing} 
\label{s:preprocessing}
While not much can be done for ads that were not flagged as political, we tried to exclude non-political content that was incorrectly flagged as such. To do so, we \textbf{(1)} further refined the collected data by employing a \textit{keyword filtering} mechanism. Additionally, to gain insights into the political campaign strategies, we \textbf{(2)} filtered ads based on their \textit{party affiliation}. The overall pipeline is summarized in Figure~\ref{fig:overview}.

\paragraph{Keyword Filtering.} We used a snowball sampling technique, as employed in previous studies \cite{conover2012partisan, yang2019bot, aiyappa2023multi}, to develop a list of relevant keywords. Starting with seed terms ``government'' and ``election'', we iteratively expanded our keyword list by identifying frequently co-occurring terms. Two authors independently reviewed and agreed on each new keyword addition. This process resulted in a final list of 43 keywords (see Figure \ref{fig:keywords}). The observed exponential decay in keyword popularity suggests that the inclusion of additional keywords would not substantially alter the corpus structure. 
Subsequently, we used a stemmed version of these keywords as a filtering mechanism to the initial dataset (PA) to extract a subset of ads more aligned to our research focus.
This process led to the creation of a second dataset, named PAK (Political Advertisements Keywords), containing 47,503 ads, and excluding 111 sponsors. While some of these keywords are of somewhat general nature and can therefore be the cause of false positives, it is important to note that they are just an additional layer of filtering on top of what is already a dataset comprising political ads (according to Meta).

\paragraph{Party Affiliation.} To gain insights into campaign strategies of major Australian political parties, a political science expert annotated a list comprising of all potential sponsors and page names, identifying each with the appropriate party and electorate where applicable. \new{Despite expecting this to be a fairly objective annotation task, to assess reliability, we had a second political science expert independently annotate 10\% of the dataset, resulting in a Cohen’s $\kappa$ inter-annotator agreement of 0.906.} We then joined this annotated list with our original data. The resulting dataset, PAP (Political Advertisements Parties), consists of 30,767 advertisements and exclusively contains ads from candidates contesting the 2022 federal election. Importantly, PAP only includes advertisements sponsored by recognized political parties, ensuring a focused analysis of official party strategies.

\paragraph{}
We want to clarify that the two preprocessing strategies are orthogonal and provide unique insights. 
This approach allows us to conduct two levels of analysis: a broad examination of the entire political advertising ecosystem, and a targeted study of official party campaign strategies.
A summary of the 3 datasets\footnote{Code and datasets are available online at {https://doi.org/10.5281/zenodo.13661563} (code repository link is available in Zenodo's dataset description).} with their characteristics can be found in Table \ref{tab:datasets}.

\begin{table}[t]
\centering
\small
\begin{tabular}{@{}lccc@{}}
\toprule[1.5pt]
\textbf{Name} & \textbf{\# Entries} & \textbf{\% of total} & \textbf{Description} \\
\midrule
\midrule
PA & 56,224 & 100\% & All ads \\
\midrule
PAK & 47,503 & 84.5\% & Ads containing $\geq 1$ keyword \\
PAP & 30,767 & 54.7\% & Ads associated to a party \\
\bottomrule[1.5pt]
\end{tabular}
\caption{Summary of datasets derived from political advertisements on Meta platforms during the 2022 Australian federal election campaign. PA represents the full dataset, while PAK and PAP are subsets based on keyword presence and party association respectively.}
\label{tab:datasets}
\end{table}


\section{Descriptive Statistics}
\label{s:descriptive}

This section presents a general overview of the statistics of the PAK dataset obtained with keyword filtering which in turn contains nearly all of the PAP dataset ($\text{PAP} \cup \text{PAK} \simeq \text{PAK}$).

In the analyzed months leading up to the election, 47,503 ads across 1,438 different sponsors were published. These ads generated over 1.11B impressions Australia-wide, with a total ad spend of A\$18.4M. Each impression represents a single ad view, with multiple impressions possible per viewer, i.e., a single user might see the same advertisement many times.

\paragraph{Individual Ads Statistics.} 

\new{Table~\ref{tab:ad_stats} summarizes the distribution of ad-level performance. Across impressions, spending, duration, and efficiency, averages were consistently higher than medians, reflecting highly skewed distributions driven by a small number of extreme outliers. For example, while the median spend was under A\$50, one ad exceeded A\$137,500, and the top-performing ad reached 1 million impressions. These disparities highlight the uneven scale of advertising on the platform, with most ads remaining relatively modest while a few campaigns dominated exposure.
}

\begin{table}[t!]
\centering
\small
\begin{tabular}{@{}lccc@{}}
\toprule[1.5pt]
\textbf{Metric} & \textbf{Mean} & \textbf{Median} & \textbf{Max} \\
\midrule
\midrule
Impressions & 23,452 & 4,500 & 1,000,000 \\
Spend (A\$) & 387 & 49.5 & 137,500 \\
Duration (days) & 8.1 & 6 & 89 \\
Efficiency (impr./A\$) & 67.5 & 50 & 4,319 \\
\bottomrule[1.5pt]
\end{tabular}
\caption{Descriptive statistics of individual ads in the PAK dataset.}
\label{tab:ad_stats}
\end{table}

\begin{table}[t]
\centering
\small
\begin{tabular}{@{}lccc@{}}
\toprule[1.5pt]
\textbf{Pairs} & $\bm{R}$ ($\bm{R^2}$) & $\bm{\rho}$ & \textbf{Strength} \\
\midrule
\midrule
Spend -- Impressions & 0.72 (0.52) & 0.82 & Strong \\
Spend -- Duration & 0.10 (0.009) & 0.26 & Weak \\
Impressions -- Duration & 0.17 (0.029) & 0.30 & Weak \\
\bottomrule[1.5pt]
\end{tabular}
\caption{Correlation coefficients between ad-level metrics. Pearson's $R$ measures linear correlation, with $R^2$ denoting the proportion of explained variance. 
Spearman's $\rho$ measures rank-based (monotonic) correlation. 
All results remain statistically significant after Bonferroni correction ($\alpha=0.0042$).}
\label{tab:correlations}
\end{table}

\paragraph{Correlation Insights.}
Our analysis revealed strong correlations between key metrics (Table~\ref{tab:correlations}). Ad expenses strongly correlated with impression numbers, suggesting that higher spending generally led to greater reach. This large effect indicates that approximately 52\% of the variance in impressions can be explained by ad spending alone.
Ad duration also showed positive, albeit weaker, correlations with both expenses and impressions. All correlations reported in this paragraph and in $\S$\ref{ss:temporal} remain statistically significant even after applying a Bonferroni correction for multiple hypothesis testing (adjusted $\alpha$$=$$\frac{0.05}{12}$$=$$0.0042$), confirming these relationships are robust (more details in Appendix \ref{a:bonferroni}).

The consistently higher Spearman coefficients relative to Pearson's suggest that the underlying relationships are monotonic rather than strictly linear, likely reflecting the non-normal nature of the data.
While all correlations are statistically significant due to our large sample size ($n$$=$$47,503$), the effect sizes for duration-related correlations are small, with duration explaining less than 3\% of the variance in impressions and less than 1\% of the variance in expenses. This suggests that while ad duration has a statistically detectable relationship with both spending and impressions, its practical influence on these metrics is limited compared to other factors.

\begin{figure}[tbp]
    \includegraphics[ width=1\columnwidth]{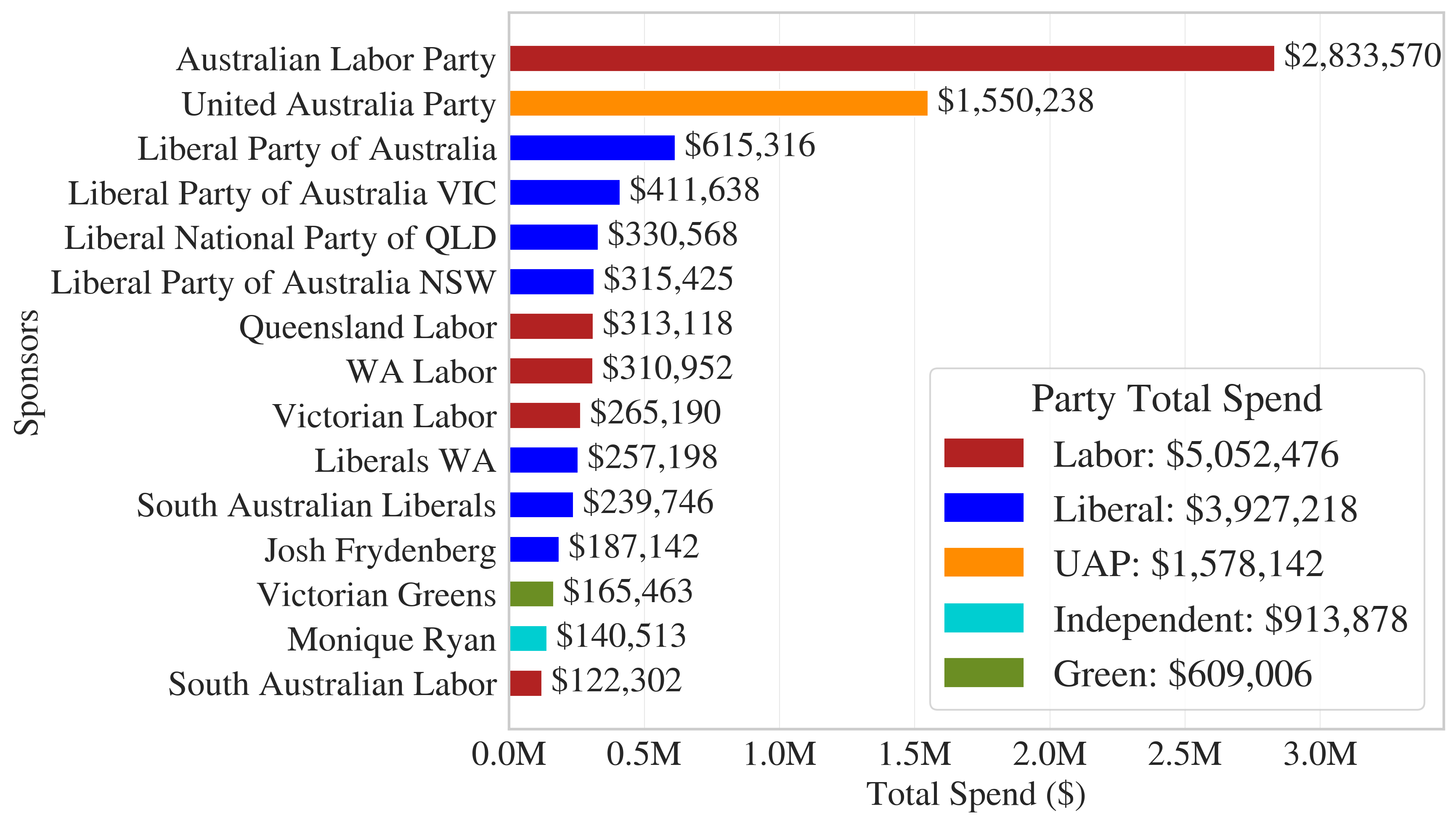}
    \caption{Total spend on political advertising by top sponsors during the 2022 election. Bars represent individual sponsors, color-coded by party affiliation. The legend shows the aggregated total spend for each major party.}
    \label{fig:funding_entities_spend}
\end{figure}

\paragraph{Sponsors Strategies.}
On average, sponsors created 33 ads each, though the median of 6 ads per sponsor indicates that most were less prolific, with the most active sponsor producing 3,239 ads. Average sponsor expenditure reached A\$12,782, but the median of A\$645 suggests that most sponsors spent relatively modest amounts, while a few big spenders significantly raised the average. The highest-spending entity, ALP, invested over A\$2.8M (Figure~\ref{fig:funding_entities_spend}). Similarly, while sponsors averaged 774k impressions, the median of 49,747 impressions again points to a skewed distribution, with the top performer achieving over 114.8 million views. Efficiency at the sponsor level achieved an average of 85.05 impressions per dollar spent (median: 70.7).

\section{Campaigning Strategies}
This analysis focuses on the five political parties in Australia,\footnote{Australian Labor Party, Liberal Coalition, United Australia Party, Idependents and Green.} examining their advertising strategies on Meta platforms during the 2022 federal election campaign. We utilized the PAP dataset, which exclusively contains ads directly sponsored by political parties, ensuring a focused analysis of official campaign strategies.

\paragraph{Spending Patterns.}
The ALP led in spending, investing over A\$5M (40.2\% of total political ad spending) across 13,047 ads in the analysed period. This surpassed the Liberal Coalition's expenditure by 28\%, with the latter spending over A\$3.9M on 9,364 ads. The UAP secured the third position, allocating A\$1.578M to their ad campaigns (\textbf{RQ1}).

Figure \ref{fig:funding_entities_spend} (legend) shows the total spend for each of the main parties. Among the top 15 party sponsors by ad spend, the ALP and Liberal Party accounted for five and seven pages respectively. The highest-ranking Australian Greens page was the Victorian Greens, placing 13th overall. Rounding out the top 15 were one page each from the UAP and an independent candidate, Monique Ryan, who contested the Kooyong electorate in Victoria taking the seat held since 2010 by Josh Frydenberg (Liberal), Treasurer and Liberal Party Deputy Leader.

\paragraph{Reach and Impressions.}
Overall, party-sponsored ads generated over 714M impressions across Australia during the campaign. As shown in Table~\ref{tab:party_stats}, ALP accounted for 40.9\% of all impressions, with the Liberal Coalition following at just over 36\%. The distribution of spending broadly mirrors the eventual vote shares, with the Liberal Coalition converting slightly fewer impressions into a higher vote share (35.7\%) compared to ALP (32.6\%). 
The Greens and the UAP attracted significantly smaller audiences. Despite outspending the Greens by more than double, the UAP generated a similar number of impressions, leading to the highest cost per impression (A\$42.2) among the major parties and translating into just 4.1\% of the vote. In contrast, the Greens achieved a lower cost per impression and converted their reach into a considerably higher vote share (12.3\%), suggesting a more engaged or better-targeted audience.

\begin{table}[t]
\centering
\scriptsize
\begin{tabular}{@{}lccccc@{}}
\toprule[1.5pt]
\textbf{Party} & \textbf{Spend} & \textbf{Ads} & \textbf{Impressions} & \textbf{CPI} & \textbf{Vote (\%)} \\
\midrule
\midrule
ALP & 5.0 A\$M & 13,047 & 292 M & 17.3 & 32.58 \\
Liberal Coalition & 3.9 A\$M & 9,364 & 261 M & 15 & 35.70 \\
UAP & 1.6 A\$M & 416 & 37.3 M & 42.2 & 4.12 \\
Independents & 0.9 A\$M & 2,844 & 51 M & 17.9 & 5.29 \\
Greens & 0.6 A\$M & 2,088 & 31.8 M & 19.1 & 12.25 \\
\bottomrule[1.5pt]
\end{tabular}
\caption{Party-level spending, impressions, and vote share. Vote percentages do not sum to 100\% due to other minor parties not shown. ALP: Australian Labor Party; UAP: United Australia Party; CPI: Cost Per Impression (A\$).}
\label{tab:party_stats}
\end{table}

\begin{figure}[t]
\includegraphics[width=1.0\columnwidth]{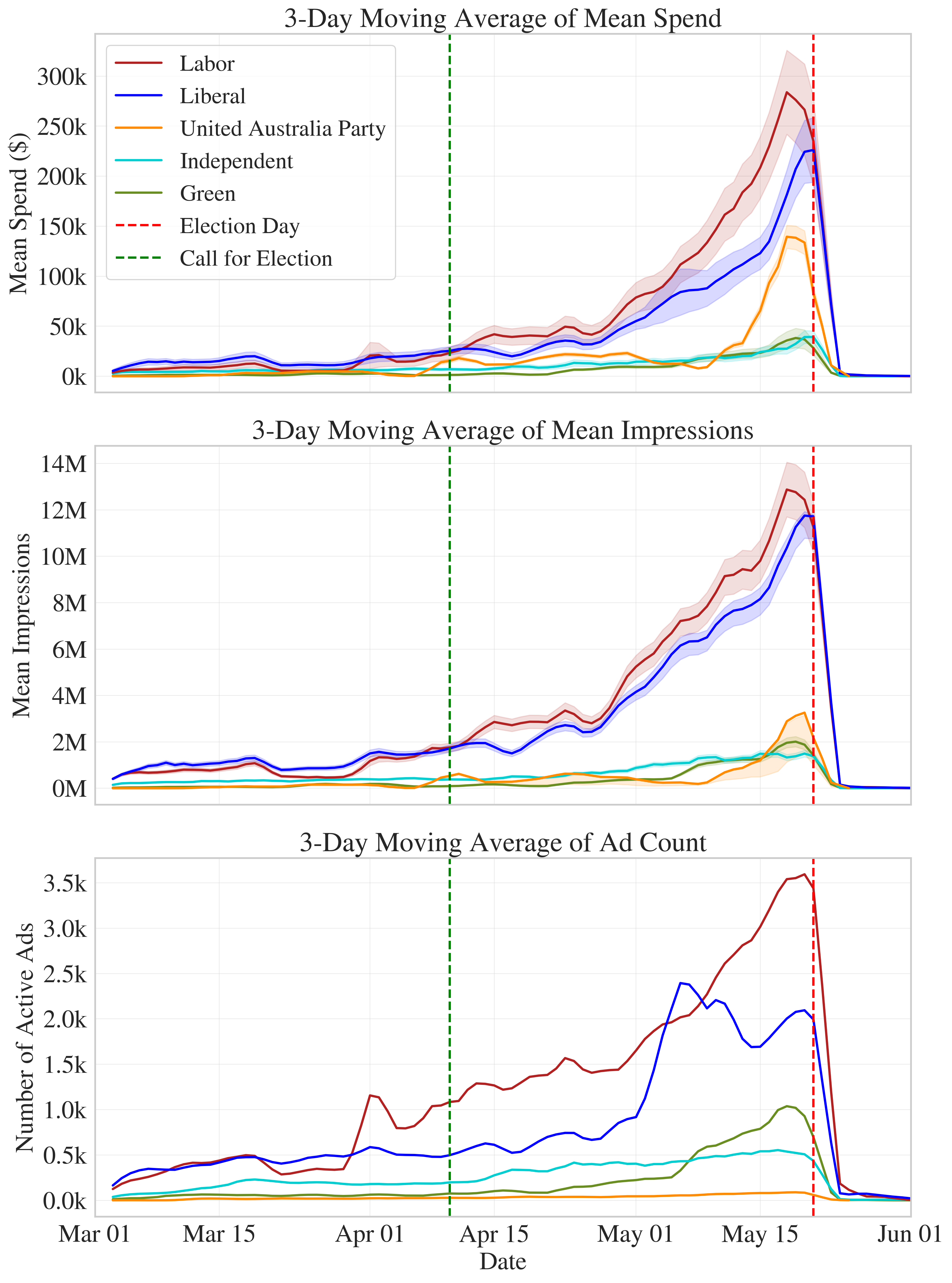}
    \caption{Time series of the total AUD spend, the number of impressions, and the number of unique ads for each day and party in the PAP dataset. In the two upper panels the dashed area represents lower and upper bounds, with the solid line corresponding to the mean value. In all plots we use a 3-day moving average.}
    \label{fig:ads_time_series}
\end{figure}

\subsection{Temporal Analysis}
\label{ss:temporal}
To understand the evolution of political advertising strategies over time (\textbf{RQ2}), we conducted a time series (TS) analysis of daily ad spend, impressions, and ad count for major Australian parties (Figure~\ref{fig:ads_time_series}).

A visual inspection reveals a notable increase in activity beginning April 10\textsuperscript{th}—the day the election was called—especially for the Labor and Liberal parties. This trend intensifies as election day approaches, with a sharp peak on May 18\textsuperscript{th}, just before the voting blackout period. 

To statistically validate this trend, we used the Mann-Kendall test, which confirmed a significant upward trend in both spend and impressions over time ($\tau$$=$$0.925$, $p$$=$$8.86\times10^{-35}$).

However, strong temporal trends can produce spurious correlations due to autocorrelation in non-stationary data. Therefore, we tested for stationarity using the KPSS test, which confirmed that all three series—spend, impressions, and ad count—were non-stationary (p-values = 0.01). 

To remove the trend and allow valid correlation analysis, we applied first-order differencing. A second KPSS test showed that the differenced series were now stationary (p-values = 0.10), validating the transformation (see Appendix~\ref{app:temporal} for details).

With stationary data, we computed correlations between the differenced series (Table \ref{tab:temporal_corr}). We found a robust relationship between changes in spend and impressions, indicating that 89\% of the variance in day-to-day changes in impressions can be explained by changes in spend. Additional strong correlations were found between differenced ad count and impressions, and between spend and ad count. Refer to Appendix \ref{a:bonferroni} for more details.
The exceptionally high correlation between spend and impressions is not only expected in digital advertising, but also likely amplified by Meta’s pricing model, where costs are often directly tied to reach metrics. This systematic relationship meant that parties could reliably predict the scale of their audience based on investment levels, enabling deliberate calibration of ad reach. Different parties exploited this dynamic in varied ways: Labor and Liberal used high ad volume, while the UAP focused on fewer but more impactful ads. Online advertising peaked just before the election, capitalizing on digital platforms' exemption from Australia’s broadcast blackout.\footnote{\url{https://www.acma.gov.au/election-and-referendum-blackout-periods}}

\begin{figure*}[ht]
    \centering
    \includegraphics[width=1\linewidth]{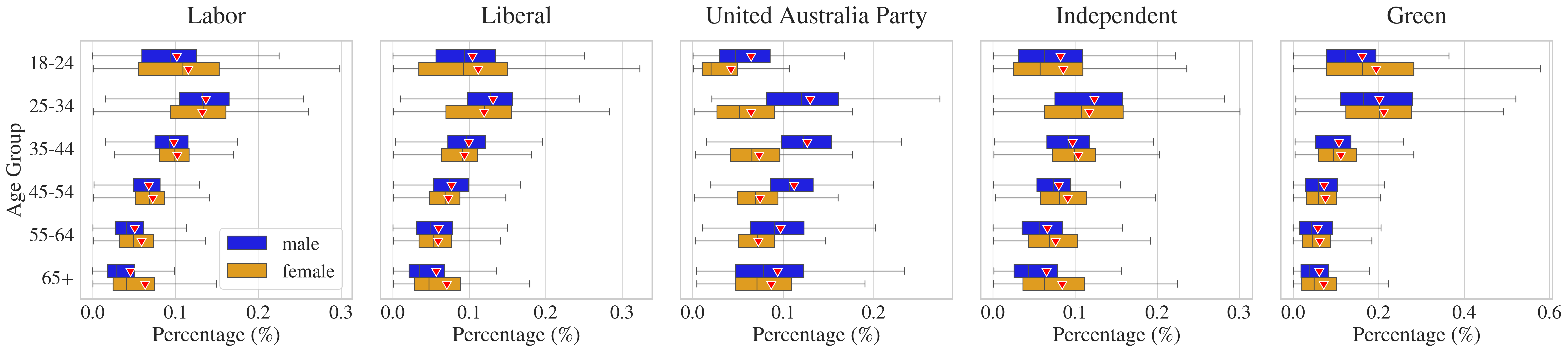}
    \caption{Distribution of ad impressions across demographic categories for major Australian political parties during the 2022 campaign. Each box plot represents the variation in the percentage of ad impressions reaching a specific age group and gender combination. The whiskers show the full range of values, boxes indicate the interquartile range, and triangles mark the mean.}
    \label{fig:demographics}
\end{figure*}

\begin{table}[t]
\centering
\small
\begin{tabular}{lccc}
\toprule[1.5pt]
\textbf{Pairs} & $\bm{R}$ ($\bm{R^2}$) & $\bm{\rho}$ & \textbf{Strength} \\
\midrule
\midrule
Spend -- Impressions & 0.94 (0.89) & 0.93 & Very strong \\
Ad Count -- Impressions & 0.83 (0.69) & 0.69 & Strong \\
Spend -- Ad Count & 0.76 (0.58) & 0.59 & Strong \\
\bottomrule[1.5pt]
\end{tabular}
\caption{Correlations between differenced time-series of daily ad metrics. Pearson's $R$ measures linear correlation, with $R^2$ denoting the proportion of explained variance. 
Spearman's $\rho$ measures rank-based (monotonic) correlation. All results remain statistically significant after Bonferroni correction ($\alpha=0.0042$).}
\label{tab:temporal_corr}
\end{table}

\begin{figure*}
    \includegraphics[width=1\linewidth]{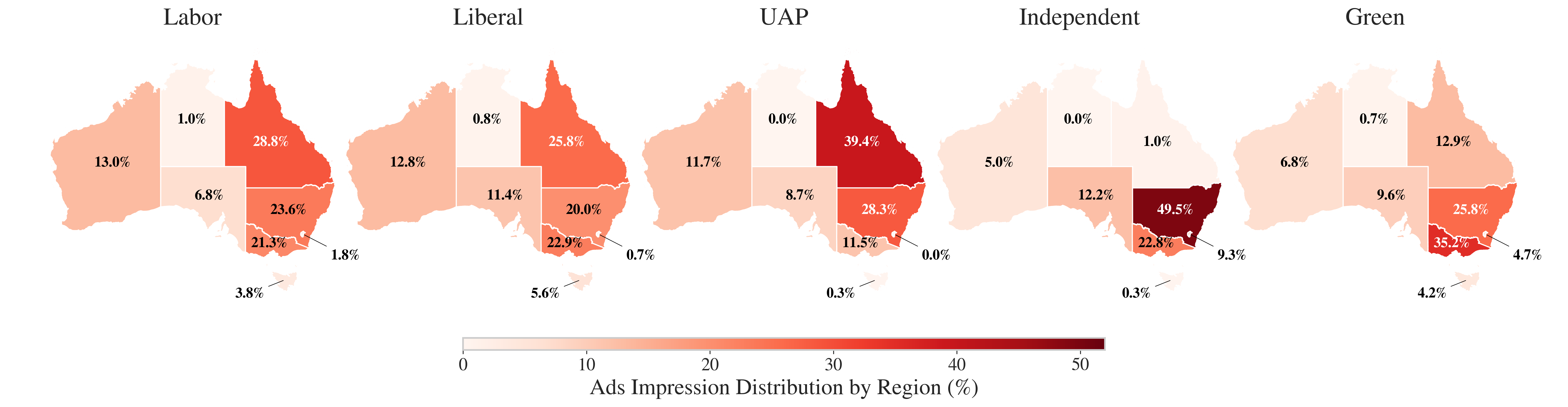}
    \caption{Regional distribution of ad impressions for major Australian political parties. Color intensity indicates the relative concentration of ad impressions in each region. We determined impressions by region multiplying the mean impressions for each ad by the proportion of ad reach attributed to each of them. The top-spend state for the Labor, Liberal, and UAP is QLD, NSW for the independents, and VIC for the Greens.}
    \label{fig:regional_distribution}
\end{figure*}

\begin{figure}[t]
    \centering
    \includegraphics[width=0.94\linewidth]{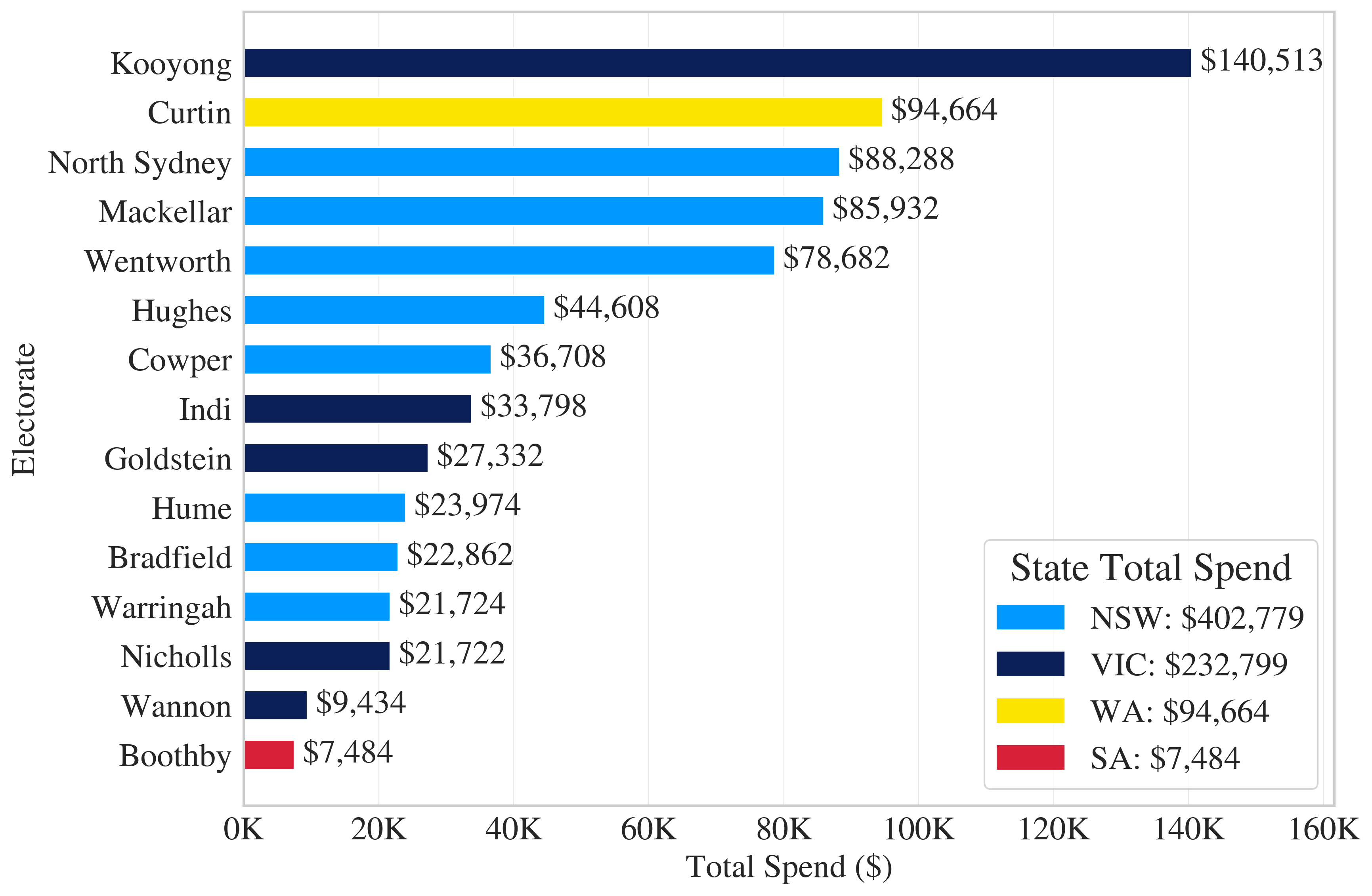}
    \caption{Total campaign spending by Independent candidates in the top 15 electorates where they ran. Bars are colored by state. The legend shows the total spend grouped by state, ranked from highest (NSW) to lowest (SA).}
    \label{fig:independents_electorates}
\end{figure}

\subsection{Demographic Analysis of Ad Impressions}
\label{ss:demographics}

Figure \ref{fig:demographics} presents how Facebook ad impressions are distributed by age and gender across major Australian political parties (\textbf{RQ2}). 

Most parties target the 18–24 and 25–34 age groups, reflecting the dominant demographics on Facebook in Australia.\footnote{https://www.statista.com/statistics/680607/australia-facebook-user-distribution-by-age-and-gender/}
The Greens particularly focus on the 18–24 group, aligning with their youth-centric mission. This emphasis on younger users could be interpreted as an attempt to engage with a demographic traditionally less involved in political processes or more susceptible to changing voting preferences. 
This targeting approach aligns with SIT \cite{tajfel1978achievement,tajfel1979integrative}, which suggests that political identities are still forming in younger demographics, making them potentially more responsive to partisan messaging that establishes "us-them" narratives as discussed in $\S$\ref{ss:politicalTheory} (\textbf{RQ3}). In contrast, the UAP distributes impressions more evenly across age groups, emphasizing older users and showing minimal engagement with the youngest demographic.

Gender-wise, the ALP tends to reach more women, especially those over 65, a trend seen in most parties. The Greens heavily target young women (18–24), while the Liberal Coalition and Independents maintain a balanced gender reach. The UAP stands out with a male-skewed audience in the 18–54 range. These gendered targeting patterns may reflect parties' understanding of how gender intersects with political identity formation, with the UAP's male-skewed approach potentially appealing to a specific social identity that aligns with their right-wing populist messaging \textbf{(RQ3)}.

\subsection{Regional Distribution}

The distribution of ad impressions across Australian regions highlights distinct strategies among political parties, as shown in Figure \ref{fig:regional_distribution} (\textbf{RQ2}). All parties show a strong focus on eastern Australia—Queensland, New South Wales, and Victoria—home to around two-thirds of the nation's population. However, the degree of this focus varies between parties.

\paragraph{Liberal \& Labor.} The major parties, Liberal Coalition and ALP, distribute their ad impressions in line with national population trends, showing a relatively balanced focus across the eastern states, with a slight tilt toward Queensland. This proportionality suggests their broader reach stems from greater financial resources, enabling high ad volume and nationwide coverage.

\paragraph{Greens \& Independents.} The Greens and Independents adopt more targeted approaches. The Greens concentrate on Victoria, likely due to limited resources and stronger urban support in Melbourne. \new{However, unlike their geographic concentration, the Greens’ messaging is thematically consistent across the country: a very large share of their advertising spend (50.03\%) is devoted to anti-mining messaging (refer to Appendix~\ref{a:mining}), reaching very high proportions in every state, peaking at 82\% in Western Australia and 81\% in Tasmania. This indicates that opposition to mining is a core focus of their campaign and is promoted nationwide rather than concentrated in one or two regions.} Independents, on the other hand, cluster ad impressions in New South Wales, as illustrated in Figure~\ref{fig:independents_electorates}, reflecting their local focus in specific electorates. Key examples include Kylea Tink (North Sydney), Sophie Scamps (Mackellar), and Allegra Spender (Wentworth), each concentrating ads in their areas.

\paragraph{United Australia Party.} UAP places heavy emphasis on Queensland, accounting for 39.4\% of their impressions—the highest regional share for any party. This reflects the party’s roots with Clive Palmer. UAP also targets New South Wales (28.3\%) and Victoria (11.5\%), while showing relatively high activity in Western Australia (11.7\%). \new{In terms of campaign themes, UAP is the party most focused on positive, mining-related advertising, devoting 4.36\% of its total spend to this topic—still a relatively small proportion overall. These mining-related ads are geographically concentrated, with New South Wales receiving the largest share relative to the party’s total spend.}

\subsection{Keywords Analysis}
\begin{figure}
    \centering
    \includegraphics[width=1\linewidth]{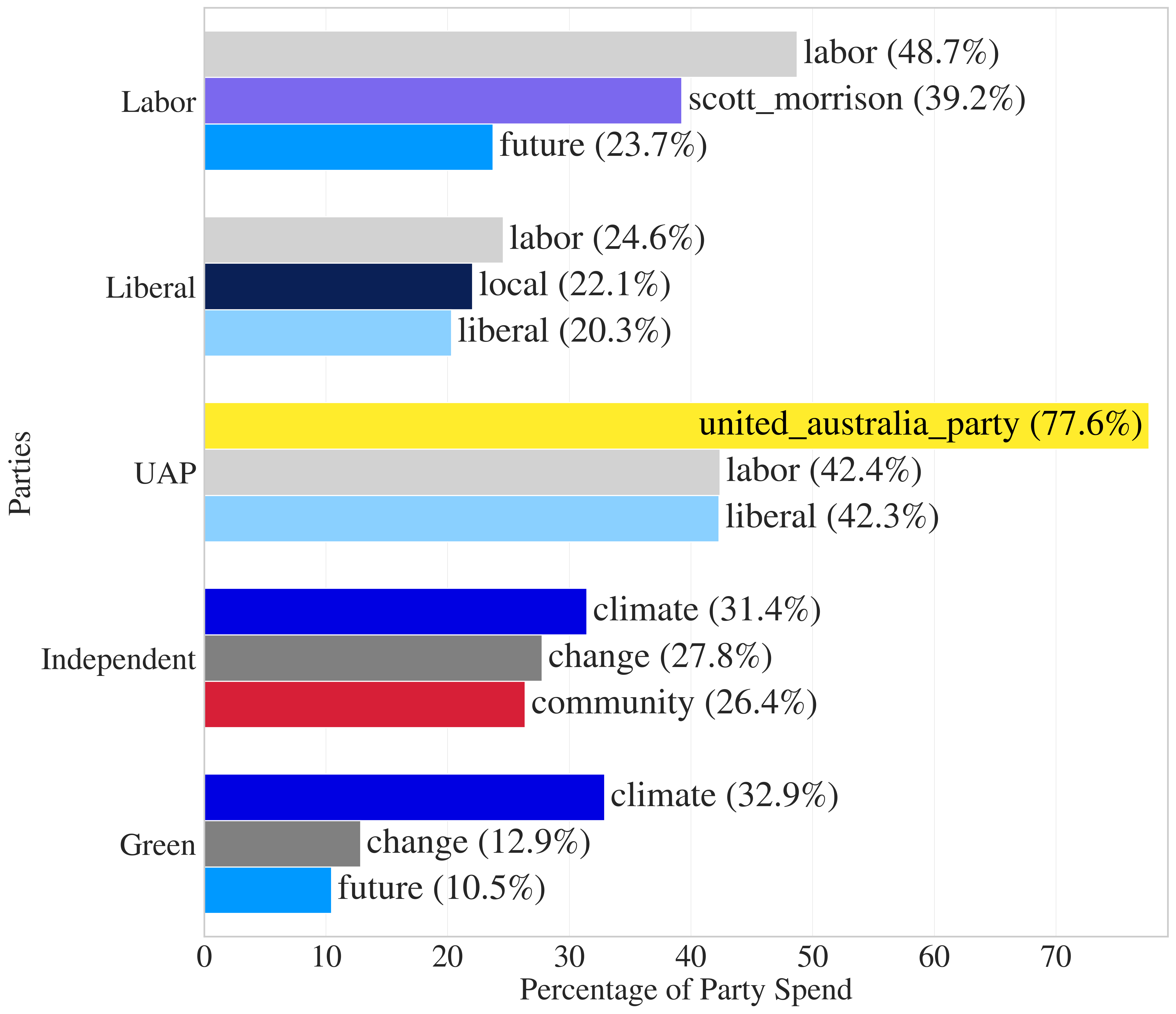}
    \caption{Most prominent keywords in each major Australian party’s ads, shown as a share of their total ad spend (common terms like “vote,” “election,” “party,” and “plan” excluded).}
    \label{fig:keyword_analysis}
\end{figure}

The analysis of keyword ad spending shows how Australian political parties used digital advertising to reinforce their ideologies and target specific voter groups (\textbf{RQ2}). As shown in Figure \ref{fig:keyword_analysis}, the Greens prioritized climate-related keywords (e.g., “climate,” “change,” “future”), aligning with their environmental agenda and appeal to younger voters, as also observed in $\S$\ref{ss:demographics}. Major parties spent heavily on promoting their own names and used contrast advertising by targeting rival party names (e.g., ALP targeting “scott\_morrison”; UAP targeting “liberal” and “labor”), reflecting a negative campaigning strategy \new{designed to reinforce in-group identity by attacking an out-group \cite{abramowitz2018negative}}. This tendency toward negative campaigning on social media platforms aligns with findings from \cite{DeckerTopic}. Independents also focused on climate-related keywords, leveraging recent climate crises to attract voters dissatisfied with major parties. This tactic of leveraging salient events to differentiate from established parties is often employed by smaller political entities to gain traction, as noted in \citet{meguid2005competition}. Overall, the keyword strategies reveal attempts to boost party visibility, attack opponents, and connect with voters on salient issues. This is especially true for smaller parties, who used issue-based appeals to form identity-based voter blocs, linking to broader political theory on tribal identity and issue alignment (\textbf{RQ3}).

\subsection{Targeting Strategies}
\label{ss:targeting}

Our initial analysis indicates that parties are possibly using targeted political advertising, focusing their campaigns in specific regions and demographics with tailored messaging. To verify these, we conduct in this section a more detailed investigation of targeting practices, drawing on methodologies established by \citet{capozzi2021clandestino} (\textbf{RQ2}).

\begin{figure}[tbp]
    \centering
    \includegraphics[width=1\linewidth]{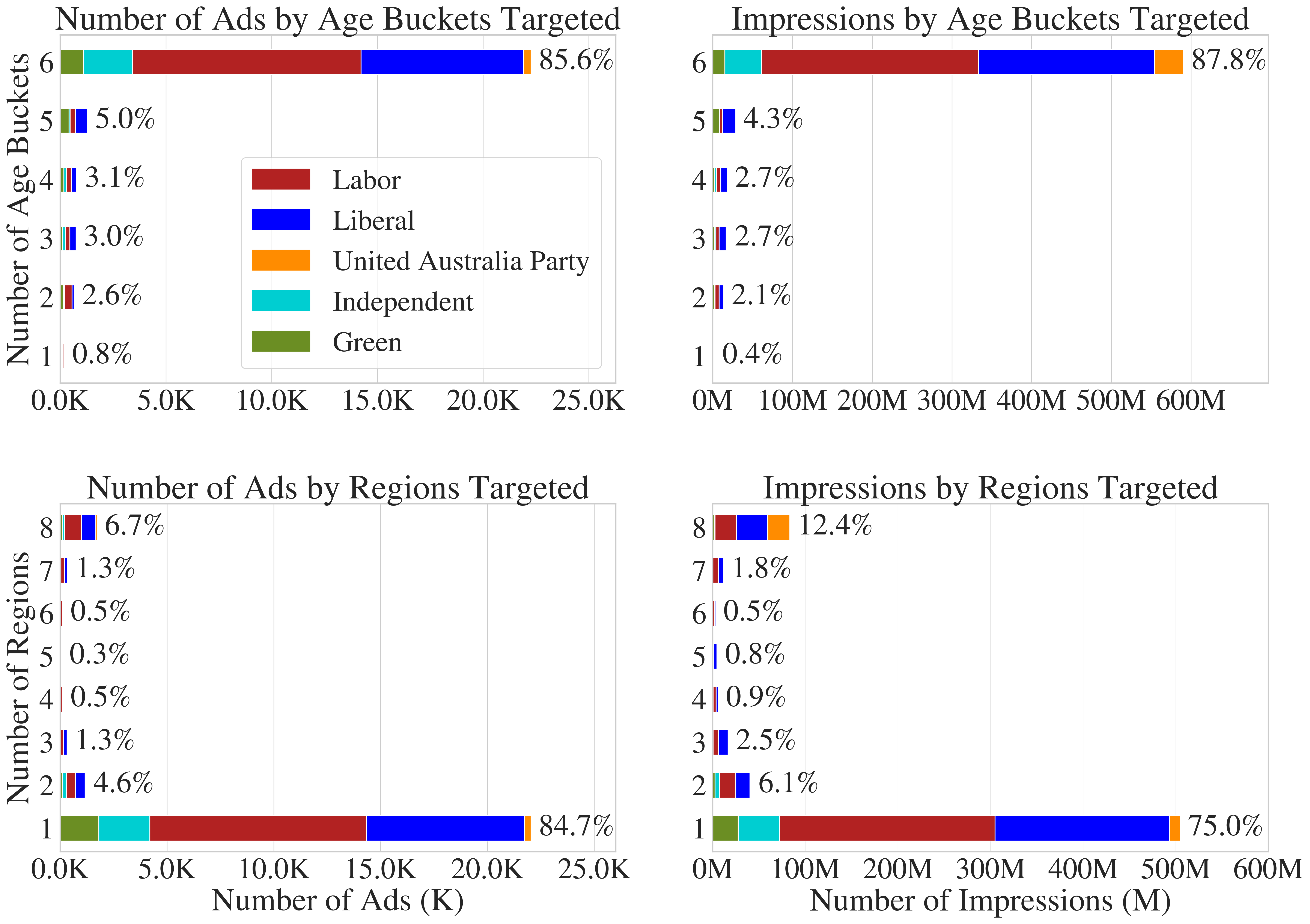}
    \caption{Distribution of ads and impressions by age buckets (top) and regions (bottom) targeted. Left panels display the number of ads, while right panels show the number of impressions. Colors represent different political parties. Percentages indicate the proportion of ads or impressions targeting a specific number of age buckets or regions.}
    \label{fig:targeting}
\end{figure}

\paragraph{Geographical Targeting.}
Geographical targeting is the most common approach, with 84.7\% of ads aimed at a single Australian region (Figure~\ref{fig:targeting} bottom-left panel). This pattern holds across all parties, with roughly 80\% of ads state-specific and 20\% broadly distributed. Surprisingly, regional and national ads have similar CPI (A\$21.4 per thousand impressions), possibly explaining the popularity of this strategy. UAP stands out with significantly higher CPI (1.9x for targeted and 2.9x for broad ads), likely due to its reliance on fewer high-spending campaigns (see $\S$\ref{ss:temporal}).

\paragraph{Age-based Targeting.}
The top-left panel in Figure~\ref{fig:targeting} reveals that the vast majority of ads (85.6\%) target all six adult age intervals (18 and above). This approach suggests a broad-reach strategy across adult age groups. We observe that these ads generate an even larger proportion of impressions (87.8\%). This indicates that parties are not extensively using age-based targeting.

\paragraph{Gender-based Targeting.}
Gender-specific targeting is even less common: only 4.8\% of ads are shown exclusively to one gender, with most of those aimed at females (62.7\%). When using a looser threshold ($\geq$ 60\% of impressions to one gender), 14.2\% of ads target females and 12.8\% target males, indicating a mild preference for female-oriented targeting. This contrasts with findings in Italy \citep{pierri2022itaelec}, where male-focused targeting is more prevalent, despite similar user demographics.

\paragraph{}
While targeting strategies—especially geographical—are evident, they reflect relatively simple use of the platform's capabilities. This may be due to limitations in the Meta API, which could obscure more advanced targeting tactics.
\new{However, this preference for broad-reach strategies is also consistent with the strategic logic of a compulsory voting system. With no need to micro-target specific segments for get-out-the-vote efforts, campaign resources are logically reallocated toward broad persuasive messaging aimed at a universal electorate \cite{hillygus2008persuadable}.}

\section{Discussion}
\label{ss:limitations}

\subsection{Summary of results}
Our analysis of political advertising on Meta platforms during the 2022 Australian federal election reveals that:
\begin{itemize}
\item \textbf{Advertising Dominance:} The ALP and Liberal Coalition dominated the digital landscape, collectively accounting for over 70\% of total ad impressions and spending. 
\item \textbf{Temporal Strategies:} Advertising activity intensified following the election call, with parties escalating their efforts as election day approached.
\item \textbf{Ads Content:} Major parties focused on self-promotion and opponent criticism, while smaller parties emphasized issue-specific messaging. Greens and Independents strategically differentiated themselves by focusing on distinct topics such as climate change and other issues neglected by major parties, appealing to voters dissatisfied with the mainstream political discourse. 
\item \textbf{Targeting Techniques:} Political entities primarily employed geographical targeting, with 84.7\% of ads targeted to a single Australian region. Age-based targeting was minimal, with 85.6\% of ads targeting all adult age groups. Gender-based targeting was limited, with only 4.8\% of ads exclusively targeting one gender. This suggests a preference for broad-reach strategies over sophisticated micro-targeting
\end{itemize}

\subsection{Rethinking On-Line Targeting in Light of Political Psychology}
Research into online campaign messaging has largely focused on who is being targeted, when, and how. While these are important empirical questions for understanding the shift from traditional to digital campaigning, less attention has been given to what persuades voters to change their minds. To address this, we can draw on longstanding research into persuasion from social and political psychology.

Humans shift between personal (“I”) and group (“We”) identities \cite{durkheim1912probleme}, a duality foundational to SIT and Self-Categorization Theory (SCT) \cite{tajfel1978achievement,tajfel1979integrative,turner1987analysis,turner1987introducing}. These theories show that persuasion is more effective (i.e., lead to enduring behaviour change) when it engages social identities \cite{mols2012makes,mols2015nudge,mols2023social}.

Emerging as a critique of rational choice models, SIT, SCT, and social constructivism \cite{berger1966social} argue that self-interest is shaped by context and identity. This lens helps explain why voters often prioritize group success over personal gain.

Applying these theories to Facebook ads from the 2022 Australian Federal Election reveals strong alignment with a ``tribal'' partisan social identity model \cite{abramowitz2018negative,campbell1960american,maggiotto1977partisan,green2004partisan,huddy2017political}.

Major parties, like ALP and the Liberal Coalition, focused on self-promotion and attacking rivals—strategies that reinforce existing identities and deepen polarization. Their investment in ads naming themselves and opponents (Figure \ref{fig:keyword_analysis}) illustrates this identity-reinforcing tactic \new{\cite{huddy2015partisan,abramowitz2018negative}}. In contrast, Independents and Greens emphasized issue-based appeals, \new{consistent with niche party theory \cite{meguid2005competition}}. Their keyword spending (31.4\% and 32.9\% respectively) suggests targeting identity groups focused on environmental concerns. This strategy creates or reinforces niche partisan affiliations beyond the major party divide.
\new{Parties likely combine identity- and issue-based appeals, reinforcing partisan identities while targeting demographics—for example, the ALP’s focus on women may reinforce partisanship while appealing to gender-specific concerns.}

This reflects a divide between party loyalists and issue-driven voters. “Rusted-on” voters maintain deep psychological attachments to parties, often inherited and resilient to policy change.\footnote{\citet{campbell1960american} defined partisanship in \textit{The American Voter} as both a set of beliefs and feelings that culminate in a sense of psychological attachment to a political party.}

Overall, we find that appeals to social identities are central to campaign strategies. However, our data cannot confirm whether such appeals change minds. Future research should go beyond correlations and use experimental designs to test causal effects. System-level factors may also shape messaging: compulsory voting in Australia might explain scare campaigns aimed at nudging the undecided, while voluntary systems may require stronger identity mobilization (e.g., MAGA in the US). These are empirical questions best addressed through established theories in social psychology and communication science.

\subsection*{Limitations.}
We acknowledge that the reliance on Meta's Ad Library data may not capture the full scope of political advertising during the election period. Some political ads may not have been properly flagged as such, leading to potential omissions in our dataset. Furthermore, our analysis does not extend to other digital platforms that may have played a role in campaign strategies. Additionally, it is crucial to consider that the apparent simplicity in targeting strategies observed may be a result of limitations in the data provided by the Meta API. \new{Some “invisible” targeting methods could interact with social identities in nuanced ways, potentially amplifying identity-based appeals in ways analysis cannot capture.
Political advertisers might be employing more advanced targeting strategies that are not visible due to restricted information disclosure.}
\new{Our analysis also focused solely on ad text, excluding images, videos, and audio components. As visual and audiovisual content often carry significant persuasive power, future work incorporating multimodal analysis could reveal additional patterns and deepen understanding of campaign strategies.}

\subsection*{Ethics Statement.}
This study used publicly available data from Meta’s Ad Library, focusing on aggregate patterns rather than individual users. No personal data was collected or analyzed. The research aims to enhance democratic transparency by examining political advertising during a federal election and follows established ethical standards, posing no direct risks. Data and code are openly shared to ensure reproducibility and encourage further research.

\bibliography{aaai22}

@book{kreiss2016prototype,
  title={Prototype politics: Technology-intensive campaigning and the data of democracy},
  author={Kreiss, Daniel},
  year={2016},
  publisher={Oxford University Press}
}

@article{dommett2019political,
  title={The political economy of Facebook advertising: Election spending, regulation and targeting online},
  author={Dommett, Katharine and Power, Sam},
  journal={The Political Quarterly},
  volume={90},
  number={2},
  year={2019},
  publisher={Wiley Online Library}
}

@article{fowler2021political,
  title={Political advertising online and offline},
  author={Fowler, Erika Franklin and Franz, Michael M and Martin, Gregory J and Peskowitz, Zachary and Ridout, Travis N},
  journal={American Political Science Review},
  volume={115},
  number={1},
  year={2021},
  publisher={Cambridge University Press}
}

@article{entous2017russian,
  title={Russian operatives used Facebook ads to exploit America's racial and religious divisions},
  author={Entous, Adam and Timberg, Craig and Dwoskin, Elizabeth},
  journal={Washington Post},
  volume={25},
  year={2017}
}

@inproceedings{capozzi2021clandestino,
  title={Clandestino or Rifugiato? Anti-immigration Facebook Ad Targeting in Italy},
  author={Capozzi, Arthur and De Francisci Morales, Gianmarco and Mejova, Yelena and Monti, Corrado and Panisson, Andr{\'e} and Paolotti, Daniela},
  booktitle={Proceedings of the 2021 CHI Conference on Human Factors in Computing Systems},
  year={2021}
}

@inproceedings{mejova2020covid,
  title={COVID-19 on Facebook ads: competing agendas around a public health crisis},
  author={Mejova, Yelena and Kalimeri, Kyriaki},
  booktitle={Proceedings of the 3rd ACM SIGCAS Conference on Computing and Sustainable Societies},
  year={2020}
}

@article{boerman2017online,
  title={Online behavioral advertising: A literature review and research agenda},
  author={Boerman, Sophie C and Kruikemeier, Sanne and Zuiderveen Borgesius, Frederik J},
  journal={Journal of Advertising},
  volume={46},
  number={3},
  year={2017},
  publisher={Taylor \& Francis}
}

@article{leerssen2021news,
  title={News from the ad archive: How journalists use the Facebook Ad Library to hold online advertising accountable},
  author={Leerssen, Paddy and Dobber, Tom and Helberger, Natali and de Vreese, Claes},
  journal={Information, Communication \& Society},
  year={2021},
  publisher={Taylor \& Francis},
  doi={10.1080/1369118X.2021.1874472},
  url={https://doi.org/10.1080/1369118X.2021.1874472}
}

@article{calvo2021global,
  title={Global spaces for local politics: An exploratory analysis of Facebook Ads in Spanish election campaigns},
  author={Calvo, Dafne and Cano-Or{\'o}n, Lorena and Baviera, Tom{\'a}s},
  journal={Social Sciences},
  volume={10},
  number={7},
  year={2021},
  publisher={MDPI},
  url={https://www.mdpi.com/2076-0760/10/7/271}
}

@article{ridout2021influence,
  title={The influence of goals and timing: How campaigns deploy Ads on Facebook},
  author={Ridout, Travis N and Fowler, Erika Franklin and Franz, Michael M},
  journal={Journal of Information Technology \& Politics},
  volume={18},
  number={3},
  year={2021},
  publisher={Taylor \& Francis}
}

@article{brodnax2022home,
  title={From Home Base to Swing States: The Evolution of Digital Advertising Strategies during the 2020 US Presidential Primary},
  author={Brodnax, NaLette M and Sapiezynski, Piotr},
  journal={Political Research Quarterly},
  volume={75},
  number={2},
  year={2022},
  publisher={SAGE Publications Sage CA: Los Angeles, CA}
}

@inproceedings{complexity_political_ads_detection,
author = {Sosnovik, Vera and Goga, Oana},
title = {Understanding the Complexity of Detecting Political Ads},
year = {2021},
isbn = {9781450383127},
publisher = {Association for Computing Machinery},
address = {New York, NY, USA},
url = {https://doi.org/10.1145/3442381.3450049},
doi = {10.1145/3442381.3450049},
booktitle = {Proceedings of the Web Conference 2021},
numpages = {12},
location = {Ljubljana, Slovenia},
series = {WWW '21}
}

@article{conover2012partisan,
  title={Partisan asymmetries in online political activity},
  author={Conover, Michael D and Gon{\c{c}}alves, Bruno and Flammini, Alessandro and Menczer, Filippo},
  journal={EPJ Data science},
  volume={1},
  year={2012},
  publisher={Springer}
}

@inproceedings{yang2019bot,
  title={Bot electioneering volume: Visualizing social bot activity during elections},
  author={Yang, Kai-Cheng and Hui, Pik-Mai and Menczer, Filippo},
  booktitle={Companion Proceedings of The 2019 World Wide Web Conference},
  year={2019}
}

@inproceedings{pierri2022itaelec,
author = {Pierri, Francesco},
title = {Political advertisement on Facebook and Instagram in the run up to 2022 Italian general election},
year = {2023},
isbn = {9798400700897},
publisher = {Association for Computing Machinery},
address = {New York, NY, USA},
url = {https://doi.org/10.1145/3578503.3583598},
doi = {10.1145/3578503.3583598},
booktitle = {Proceedings of the 15th ACM Web Science Conference 2023},
numpages = {10},
location = {Austin, TX, USA},
series = {WebSci '23}
}

@article{DeckerTopic,
	author = {Hannah Decker and Daniel Angus and Axel Bruns and Ehsan Dehghan and Phoebe Matich and Jane Tan and Laura Vodden},
	title = {Topic Diversity in Social Media Campaigning: A Study of the 2022 Australian Federal Election},
	journal = {Politics and Governance},
	volume = {12},
	number = {0},
	year = {2024},
	issn = {2183-2463},	doi = {10.17645/pag.8155},
	url = {https://www.cogitatiopress.com/politicsandgovernance/article/view/8155}
}

@article{bruns2018social,
  title={Social media in Australian federal elections: Comparing the 2013 and 2016 campaigns},
  author={Bruns, Axel and Moon, Brenda},
  journal={Journalism \& Mass Communication Quarterly},
  volume={95},
  number={2},
  year={2018},
  publisher={SAGE Publications Sage CA: Los Angeles, CA}
}

@article{meguid2005competition,
  title={Competition between unequals: The role of mainstream party strategy in niche party success},
  author={Meguid, Bonnie M},
  journal={American political science review},
  volume={99},
  number={3},
  year={2005},
  publisher={Cambridge University Press}
}

@inproceedings{aiyappa2023multi,
  title={A multi-platform collection of social media posts about the 2022 US midterm elections},
  author={Aiyappa, Rachith and DeVerna, Matthew R and Pote, Manita and Truong, Bao Tran and Zhao, Wanying and Axelrod, David and Pessianzadeh, Aria and Kachwala, Zoher and Kim, Munjung and Seckin, Ozgur Can and others},
  booktitle={Proceedings of the international AAAI conference on web and social media},
  volume={17},
  year={2023}
}

@book{simon1957models,
  title={Models of man: social and rational; mathematical essays on rational human behavior in society setting},
  author={Simon, Herbert Alexander},
  year={1957},
  publisher={Wiley}
}

@misc{tajfel1978achievement,
  title={The achievement of inter-group differentiation. Differentiation between social groups},
  author={Tajfel, Henri},
  year={1978},
  publisher={London: Academic Press London}
}

@article{tajfel1979integrative,
  title={An Integrative theory of inter-group Conflict (In WG Austin and S. Worchel, The social psychology of inter-group relations, Monterey, CA: Brooks)},
  author={Tajfel, H and Turner, JC},
  year={1979}
}

@article{berger1966social,
  title={The social construction of reality},
  author={Berger Peter, L and Luckmann, Thomas},
  journal={A Treatise in the Sociology of Knowledge},
  year={1966},
  publisher={Doubleday Garden City}
}

@article{durkheim1912probleme,
  title={Le probl{`e}me religieux et la dualit{'e} de la nature humaine},
  author={Durkheim, Emile},
  year={1912}
}

@inproceedings{o2010tweets,
  title={From tweets to polls: Linking text sentiment to public opinion time series},
  author={O'Connor, Brendan and Balasubramanyan, Ramnath and Routledge, Bryan and Smith, Noah},
  booktitle={Proceedings of the international AAAI conference on web and social media},
  volume={4},
  number={1},
  year={2010}
}

@article{tumasjan2011election,
  title={Election forecasts with Twitter: How 140 characters reflect the political landscape},
  author={Tumasjan, Andranik and Sprenger, Timm O and Sandner, Philipp G and Welpe, Isabell M},
  journal={Social science computer review},
  volume={29},
  number={4},
  year={2011},
  publisher={Sage Publications Sage CA: Los Angeles, CA}
}

@article{gayo2012no,
  title={No, you cannot predict elections with Twitter},
  author={Gayo-Avello, Daniel},
  journal={IEEE Internet Computing},
  volume={16},
  number={6},
  year={2012},
  publisher={IEEE}
}

@article{gayo2012wanted,
  title={"I Wanted to Predict Elections with Twitter and all I got was this Lousy Paper"--A Balanced Survey on Election Prediction using Twitter Data},
  author={Gayo-Avello, Daniel},
  journal={arXiv preprint arXiv:1204.6441},
  year={2012}
}

@inproceedings{metaxas2011not,
  title={How (not) to predict elections},
  author={Metaxas, Panagiotis T and Mustafaraj, Eni and Gayo-Avello, Dani},
  booktitle={2011 ieee third international conference on privacy, security, risk and trust and 2011 ieee third international conference on social computing},
  year={2011},
  organization={IEEE}
}

@article{jungherr2016twitter,
  title={Twitter use in election campaigns: A systematic literature review},
  author={Jungherr, Andreas},
  journal={Journal of information technology \& politics},
  volume={13},
  number={1},
  year={2016},
  publisher={Taylor \& Francis}
}

@inproceedings{skoric2012tweets,
  title={Tweets and votes: A study of the 2011 singapore general election},
  author={Skoric, Marko and Poor, Nathaniel and Achananuparp, Palakorn and Lim, Ee-Peng and Jiang, Jing},
  booktitle={2012 45th hawaii international conference on system sciences},
  year={2012},
  organization={IEEE}
}

@article{abramowitz2018negative,
  title={Negative partisanship: Why Americans dislike parties but behave like rabid partisans},
  author={Abramowitz, Alan I and Webster, Steven W},
  journal={Political Psychology},
  volume={39},
  year={2018},
  publisher={Wiley Online Library}
}

@book{campbell1960american,
  title={The american voter},
  author={Campbell, Angus and Converse, Philip E and Miller, Warren E and Stokes, Donald E},
  year={1960},
  publisher={John Wiley \& Sons}
}

@article{fiorina1981retrospective,
  title={Retrospective voting in American national elections},
  author={Fiorina, Morris P},
  publisher={Yale University Press},
  year={1981}
}

@book{green2004partisan,
  title={Partisan hearts and minds: Political parties and the social identities of voters},
  author={Green, Donald P and Palmquist, Bradley and Schickler, Eric},
  year={2004},
  publisher={Yale University Press}
}

@incollection{huddy2017political,
  title={Political partisanship as a social identity},
  author={Huddy, Leonie and Bankert, Alexa},
  booktitle={Oxford research encyclopedia of politics},
  year={2017}
}

@article{maggiotto1977partisan,
  title={Partisan identification and electoral choice: The hostility hypothesis},
  author={Maggiotto, Michael A and Piereson, James E},
  journal={American Journal of Political Science},
  year={1977},
  publisher={JSTOR}
}

@article{mols2012makes,
  title={What makes a frame persuasive? Lessons from social identity theory},
  author={Mols, Frank},
  journal={Evidence \& Policy},
  volume={8},
  number={3},
  year={2012},
  publisher={Policy Press}
}

@article{mols2015nudge,
  title={Why a nudge is not enough: A social identity critique of governance by stealth},
  author={Mols, Frank and Haslam, S Alexander and Jetten, Jolanda and Steffens, Niklas K},
  journal={European Journal of Political Research},
  volume={54},
  number={1},
  year={2015},
  publisher={Wiley Online Library}
}

@article{orr2020policy,
  title={The policy basis of measured partisan animosity in the United States},
  author={Orr, Lilla V and Huber, Gregory A},
  journal={American Journal of Political Science},
  volume={64},
  number={3},
  year={2020},
  publisher={Wiley Online Library}
}

@article{west2022partisanship,
  title={Partisanship as a social identity: Implications for polarization},
  author={West, Emily A and Iyengar, Shanto},
  journal={Political Behavior},
  volume={44},
  number={2},
  year={2022},
  publisher={Springer}
}

@article{turner1987analysis,
  title={The analysis of social influence},
  author={Turner, John C},
  journal={Rediscovering the social group: A self-categorization theory},
  year={1987}
}

@misc{turner1987introducing,
  title={Introducing the problem: Individual and group. Rediscovering the social group: A self-categorization theory (pp. 1--18)},
  author={Turner, JC},
  year={1987},
  publisher={Blackwell}
}

@article{kwiatkowski1992testing,
  title={Testing the null hypothesis of stationarity against the alternative of a unit root: How sure are we that economic time series have a unit root?},
  author={Kwiatkowski, Denis and Phillips, Peter CB and Schmidt, Peter and Shin, Yongcheol},
  journal={Journal of econometrics},
  volume={54},
  number={1-3},
  year={1992},
  publisher={Elsevier}
}

@article{mols2023social,
  title={The social identity approach to political leadership},
  author={Mols, Frank and Haslam, S Alexander and Platow, Michael J and Reicher, Stephen D and Steffens, Niklas K},
  year={2023}
}

@book{arya2022political,
  title={Political advertising on social media platforms during the 2022 federal election},
  author={Arya, Prachi},
  year={2022},
  publisher={Australia Institute}
}

@inproceedings{islam2023weakly,
  title={Weakly supervised learning for analyzing political campaigns on facebook},
  author={Islam, Tunazzina and Roy, Shamik and Goldwasser, Dan},
  booktitle={Proceedings of the International AAAI Conference on Web and Social Media},
  volume={17},
  year={2023}
}

@inproceedings{bouchaud2024beyond,
  title={Beyond the Guidelines: Assessing Meta's Political Ad Moderation in the EU},
  author={Bouchaud, Paul and Li{\'e}nard, Jean F},
  booktitle={Proceedings of the 2024 ACM on Internet Measurement Conference},
  year={2024}
}

@article{bar2024systematic,
  title={Systematic discrepancies in the delivery of political ads on Facebook and Instagram},
  author={B{\"a}r, Dominik and Pierri, Francesco and De Francisci Morales, Gianmarco and Feuerriegel, Stefan},
  journal={PNAS nexus},
  volume={3},
  number={7},
  year={2024},
  publisher={Oxford University Press US}
}

@inproceedings{islam2023climate,
  title={Analysis of climate campaigns on social media using bayesian model averaging},
  author={Islam, Tunazzina and Zhang, Ruqi and Goldwasser, Dan},
  booktitle={Proceedings of the 2023 AAAI/ACM Conference on AI, Ethics, and Society},
  year={2023}
}

@inproceedings{islam2022covid,
  title={Understanding covid-19 vaccine campaign on facebook using minimal supervision},
  author={Islam, Tunazzina and Goldwasser, Dan},
  booktitle={2022 IEEE International Conference on Big Data (Big Data)},
  year={2022},
  organization={IEEE}
}

@article{oprea2024moving,
  title={Moving toward the median: Compulsory voting and political polarization},
  author={Oprea, Alexandra and Martin, Lucy and Brennan, Geoffrey H},
  journal={American Political Science Review},
  volume={118},
  number={4},
  year={2024},
  publisher={Cambridge University Press}
}

@incollection{Birch2009compulsory,
    author = {Birch, Sarah},
    isbn = {9780719077623},
    title = {59Compulsory voting and election campaigns},
    booktitle = {Full Participation: A Comparative Study of Compulsory Voting},
    publisher = {Manchester University Press},
    year = {2009},
    month = {02},
    doi = {10.7228/manchester/9780719077623.003.0004},
    url = {https://doi.org/10.7228/manchester/9780719077623.003.0004},
    eprint = {https://academic.oup.com/manchester-scholarship-online/book/0/chapter/173824811/chapter-ag-pdf/44755122/book_16712_section_173824811.ag.pdf},
}

@article{jackman2001compulsory,
  title={Compulsory voting},
  author={Jackman, Simon},
  year = {2001},
}

@misc{hillygus2008persuadable,
  title={The Persuadable Voter: Wedge Issues in Presidential Campaigns},
  author={Hillygus, D Sunshine and Shields, Todd G},
  year={2008},
  publisher={Princeton University Press}
}

@article{singh2013compulsory,
  title={Compulsory voting and the dynamics of partisan identification},
  author={Singh, Shane and Thornton, Judd},
  journal={European Journal of Political Research},
  volume={52},
  number={2},
  year={2013},
  publisher={Wiley Online Library}
}

@article{huddy2015partisan,
  title={Expressive partisanship: Campaign involvement, political emotion, and partisan identity},
  author={Huddy, Leonie and Mason, Lilliana and Aar{\o}e, Lene},
  journal={American Political Science Review},
  volume={109},
  number={1},
  year={2015},
  publisher={Cambridge University Press}
}

@incollection{bankert2020origins,
  title={The origins and effect of negative partisanship},
  author={Bankert, Alexa},
  booktitle={Research handbook on political partisanship},
  year={2020},
  publisher={Edward Elgar Publishing}
}

@article{garrett2020moral,
  title={The moral roots of partisan division: How moral conviction heightens affective polarization},
  author={Garrett, Kristin N and Bankert, Alexa},
  journal={British Journal of Political Science},
  volume={50},
  number={2},
  year={2020},
  publisher={Cambridge University Press}
}

@misc{meguellati2025,
      title={Towards Detecting Persuasion on Social Media: From Model Development to Insights on Persuasion Strategies}, 
      author={Elyas Meguellati and Stefano Civelli and Pietro Bernardelle and Shazia Sadiq and Irwin King and Gianluca Demartini},
      year={2025},
      eprint={2503.13844},
      archivePrefix={arXiv},
      primaryClass={cs.CL},
      url={https://arxiv.org/abs/2503.13844}, 
}

\newpage
\section*{Ethics Checklist}

\begin{enumerate}

\item For most authors...
\begin{enumerate}
    \item  Would answering this research question advance science without violating social contracts, such as violating privacy norms, perpetuating unfair profiling, exacerbating the socio-economic divide, or implying disrespect to societies or cultures?
    \answerYes{Yes.}
  \item Do your main claims in the abstract and introduction accurately reflect the paper's contributions and scope?
    \answerYes{Yes.}
   \item Do you clarify how the proposed methodological approach is appropriate for the claims made? 
    \answerYes{Yes.}
   \item Do you clarify what are possible artifacts in the data used, given population-specific distributions?
    \answerYes{Yes. See $\S$\ref{ss:limitations}.}
  \item Did you describe the limitations of your work?
    \answerYes{Yes. See $\S$\ref{ss:limitations}.}
  \item Did you discuss any potential negative societal impacts of your work?
    \answerNA{N/A}
      \item Did you discuss any potential misuse of your work?
    \answerNA{N/A}
    \item Did you describe steps taken to prevent or mitigate potential negative outcomes of the research, such as data and model documentation, data anonymization, responsible release, access control, and the reproducibility of findings?
    \answerYes{Yes.}
  \item Have you read the ethics review guidelines and ensured that your paper conforms to them?
    \answerYes{Yes.}
\end{enumerate}

\item Additionally, if your study involves hypotheses testing...
\begin{enumerate}
  \item Did you clearly state the assumptions underlying all theoretical results?
    \answerNA{N/A}
  \item Have you provided justifications for all theoretical results?
    \answerNA{N/A}
  \item Did you discuss competing hypotheses or theories that might challenge or complement your theoretical results?
    \answerNA{N/A}
  \item Have you considered alternative mechanisms or explanations that might account for the same outcomes observed in your study?
    \answerNA{N/A}
  \item Did you address potential biases or limitations in your theoretical framework?
    \answerNA{N/A}
  \item Have you related your theoretical results to the existing literature in social science?
    \answerNA{N/A}
  \item Did you discuss the implications of your theoretical results for policy, practice, or further research in the social science domain?
    \answerNA{N/A}
\end{enumerate}

\item Additionally, if you are including theoretical proofs...
\begin{enumerate}
  \item Did you state the full set of assumptions of all theoretical results?
    \answerNA{N/A}
	\item Did you include complete proofs of all theoretical results?
    \answerNA{N/A}
\end{enumerate}

\item Additionally, if you ran machine learning experiments...
\begin{enumerate}
  \item Did you include the code, data, and instructions needed to reproduce the main experimental results (either in the supplemental material or as a URL)?
    \answerNA{N/A}
  \item Did you specify all the training details (e.g., data splits, hyperparameters, how they were chosen)?
    \answerNA{N/A}
     \item Did you report error bars (e.g., with respect to the random seed after running experiments multiple times)?
    \answerNA{N/A}
	\item Did you include the total amount of compute and the type of resources used (e.g., type of GPUs, internal cluster, or cloud provider)?
    \answerNA{N/A}
     \item Do you justify how the proposed evaluation is sufficient and appropriate to the claims made? 
    \answerNA{N/A}
     \item Do you discuss what is ``the cost`` of misclassification and fault (in)tolerance?
    \answerNA{N/A}
  
\end{enumerate}

\item Additionally, if you are using existing assets (e.g., code, data, models) or curating/releasing new assets, \textbf{without compromising anonymity}...
\begin{enumerate}
  \item If your work uses existing assets, did you cite the creators?
    \answerNA{N/A}
  \item Did you mention the license of the assets?
    \answerNA{N/A} 
  \item Did you include any new assets in the supplemental material or as a URL?
    \answerYes{Yes. See $\S$\ref{s:data_curation} in which we provide code and datasets.}
  \item Did you discuss whether and how consent was obtained from people whose data you're using/curating?
    \answerNA{N/A}
  \item Did you discuss whether the data you are using/curating contains personally identifiable information or offensive content?
    \answerNA{N/A}
\item If you are curating or releasing new datasets, did you discuss how you intend to make your datasets FAIR?
\answerYes{Yes. In $\S$\ref{s:preprocessing}, we link to a Zenodo dataset repository, as per FAIR guidelines.}
\item If you are curating or releasing new datasets, did you create a Datasheet for the Dataset? 
\answerYes{Yes. The  Zenodo  repository  includes  a  detailed Datasheet file where we addressed the required questions.}
\end{enumerate}

\item Additionally, if you used crowdsourcing or conducted research with human subjects, \textbf{without compromising anonymity}...
\begin{enumerate}
  \item Did you include the full text of instructions given to participants and screenshots?
    \answerNA{N/A}
  \item Did you describe any potential participant risks, with mentions of Institutional Review Board (IRB) approvals?
    \answerNA{N/A}
  \item Did you include the estimated hourly wage paid to participants and the total amount spent on participant compensation?
    \answerNA{N/A}
   \item Did you discuss how data is stored, shared, and deidentified?
   \answerNA{N/A}
\end{enumerate}

\end{enumerate}

\newpage

\appendix

\section{Statistical Details for Temporal Analysis}
\label{app:temporal}
\subsection{Stationarity Testing}
To assess the validity of correlation metrics in the presence of trending data, we applied the KPSS test to our three series: daily ad count, spend, and impressions. All were non-stationary in their original form:
\begin{itemize}
    \item Ad Count: KPSS = 1.37, $p = 0.01$
    \item Spend: KPSS = 1.08, $p = 0.01$
    \item Impressions: KPSS = 1.21, $p = 0.01$
\end{itemize}
After first-order differencing, all series became stationary:
\begin{itemize}
    \item Differenced Ad Count: KPSS = 0.12, $p = 0.10$
    \item Differenced Spend: KPSS = 0.16, $p = 0.10$
    \item Differenced Impressions: KPSS = 0.16, $p = 0.10$
\end{itemize}

It is important to note that the p-values reported by the KPSS test are limited to the values 0.01 and 0.10. This is because the implementation in statsmodels interpolates p-values from Table 1 in \citet{kwiatkowski1992testing} and returns boundary points when the test statistic falls outside the table of critical values. Specifically, if the p-value would be smaller than 0.01 or larger than 0.10, the function returns these boundary values instead. This explains why our non-stationary series all show $p = 0.01$ (indicating strong evidence against stationarity) and our stationary series show $p = 0.10$ (indicating no evidence against stationarity).

\subsection{Time Series Differencing}
\begin{figure}[h]
\includegraphics[width=1.0\columnwidth]{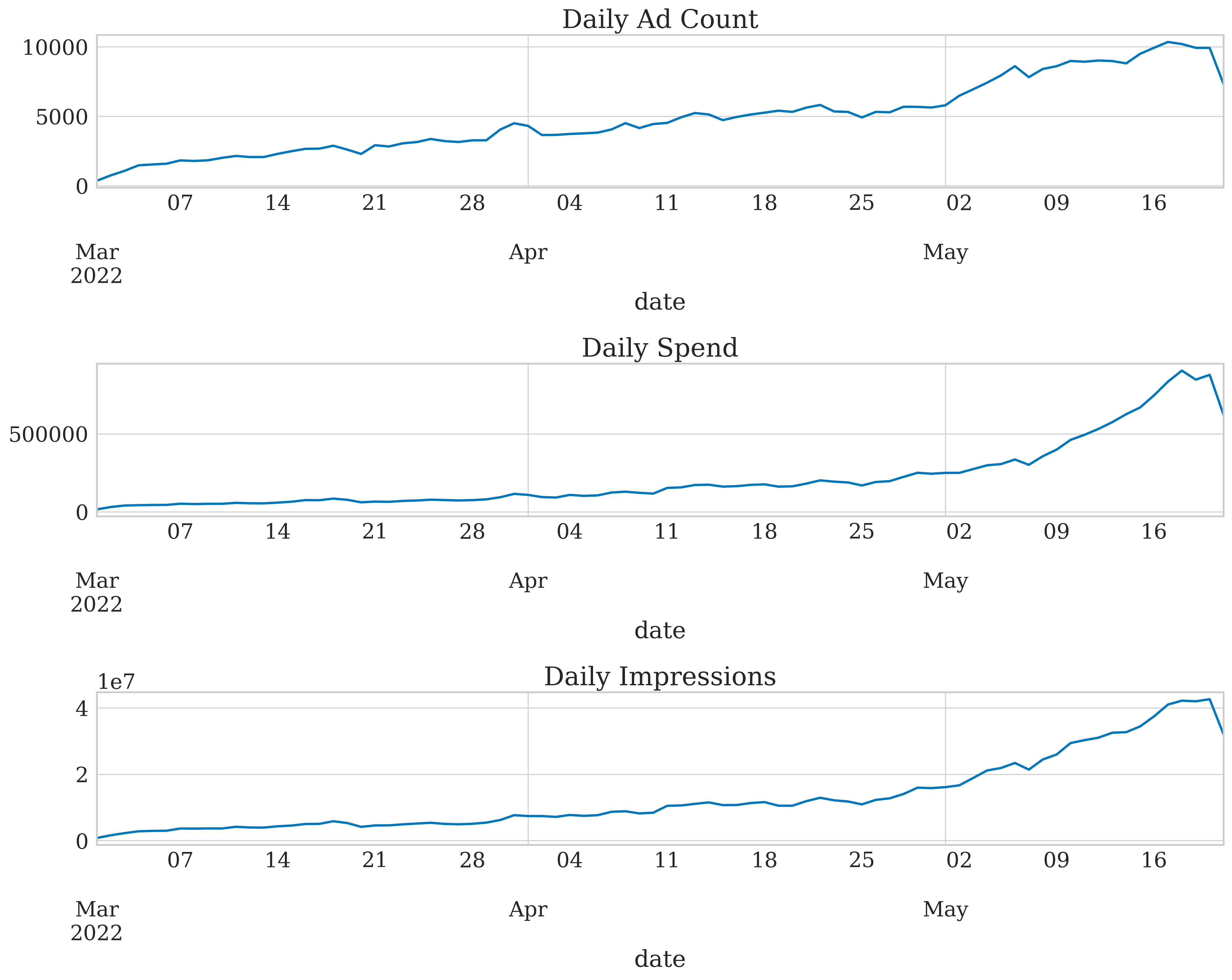}
    \caption{Original time series data showing daily ad counts, spending, and impressions from March to May 2022. All three metrics show strong upward trends, particularly accelerating in May, with a sharp decline at the end of the period coinciding with the election blackout. The daily ad count reaches approximately 10,000 at its peak, daily spend approaches 800,000 AUD, and daily impressions reach over 40 million.}
    \label{fig:original_time_series}
\end{figure}

Figure \ref{fig:original_time_series} displays the raw time series data for our three metrics. The consistent upward trend across all three metrics indicates non-stationarity, which was confirmed by our KPSS test results. We observe that all three metrics show relatively steady growth from March through April, followed by a more dramatic increase in early-to-mid May, and then a sharp drop-off near the end of the observation period corresponding to the election blackout period.

\begin{figure}[t]
\includegraphics[width=1.0\columnwidth]{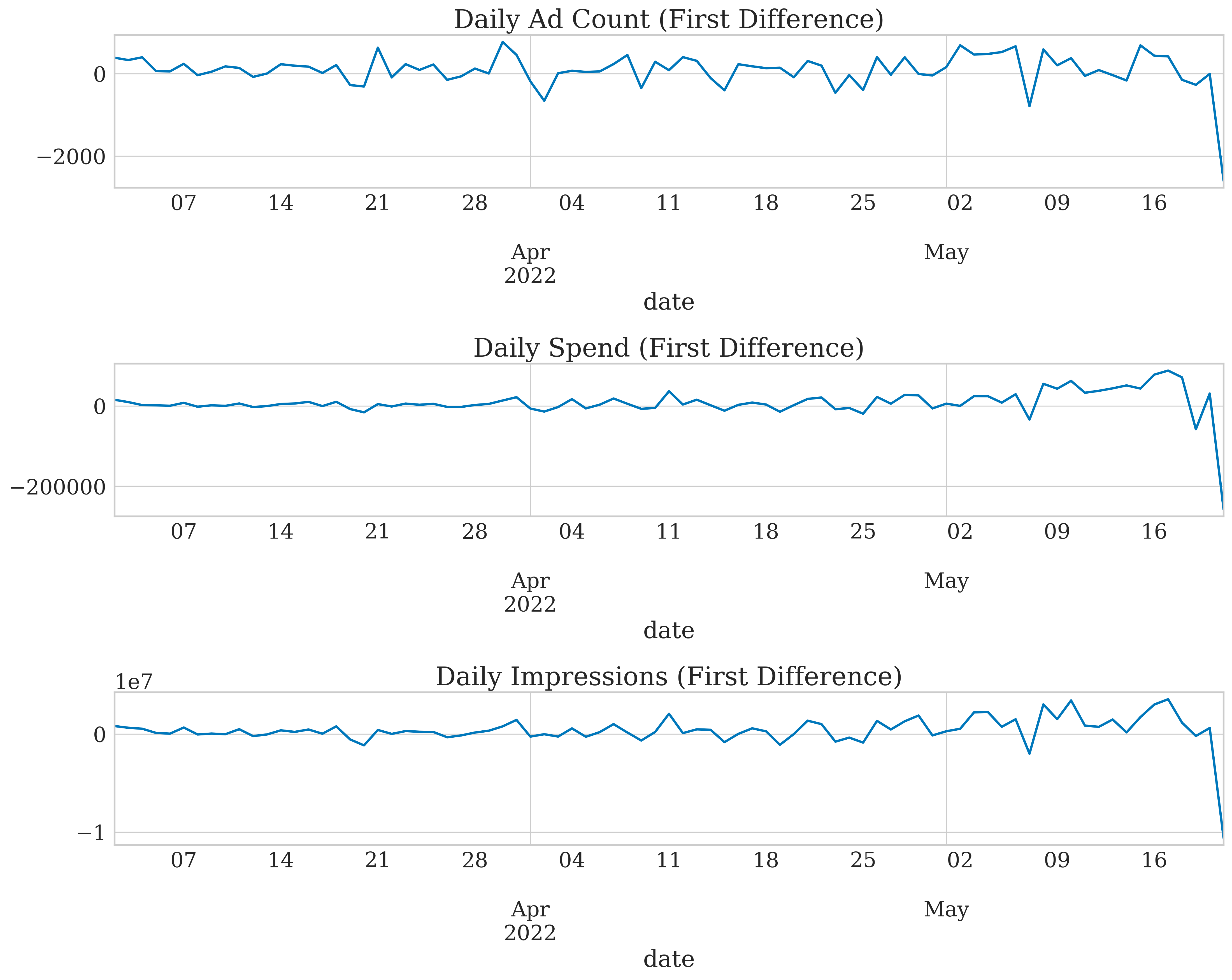}
    \caption{First-differenced time series showing daily changes in ad counts, spending, and impressions. The differencing process removes the trend component, resulting in stationary series fluctuating around zero.}
    \label{fig:differenced_time_series}
\end{figure}

Figure \ref{fig:differenced_time_series} presents the first-differenced versions of our time series, where each point represents the day-to-day change in the respective metric. The differencing procedure effectively removed the trend component, resulting in stationary series that fluctuate around zero, as confirmed by our second round of KPSS tests.

\subsection{Correlation Analysis with Differenced Data}

\begin{figure}[t]
    \centering
    \begin{subfigure}[b]{0.48\columnwidth}
        \includegraphics[width=\textwidth]{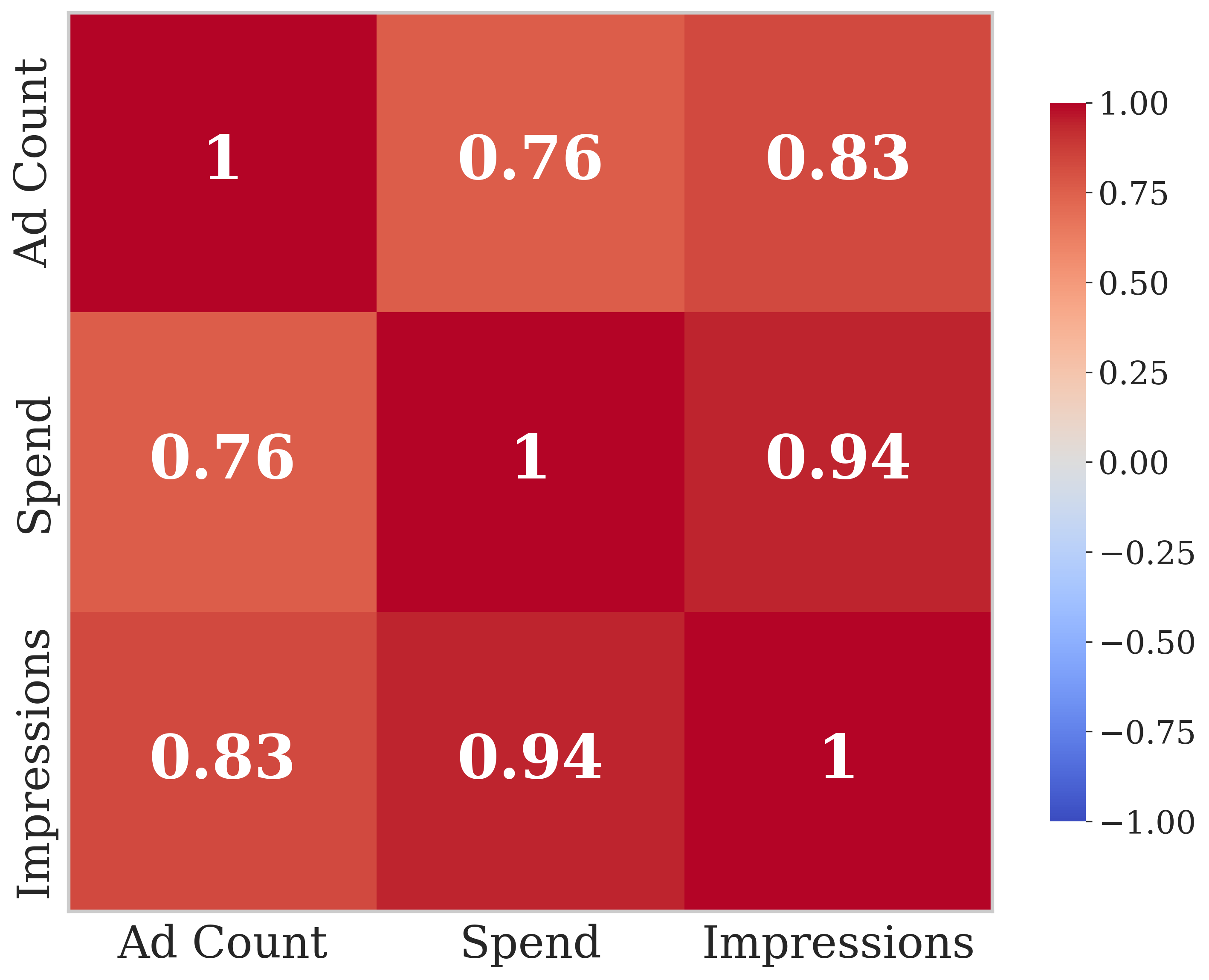}
        \caption{Pearson correlation matrix of stationary time series.}
        \label{fig:pearson_corr}
    \end{subfigure}
    \hfill
    \begin{subfigure}[b]{0.48\columnwidth}
        \includegraphics[width=\textwidth]{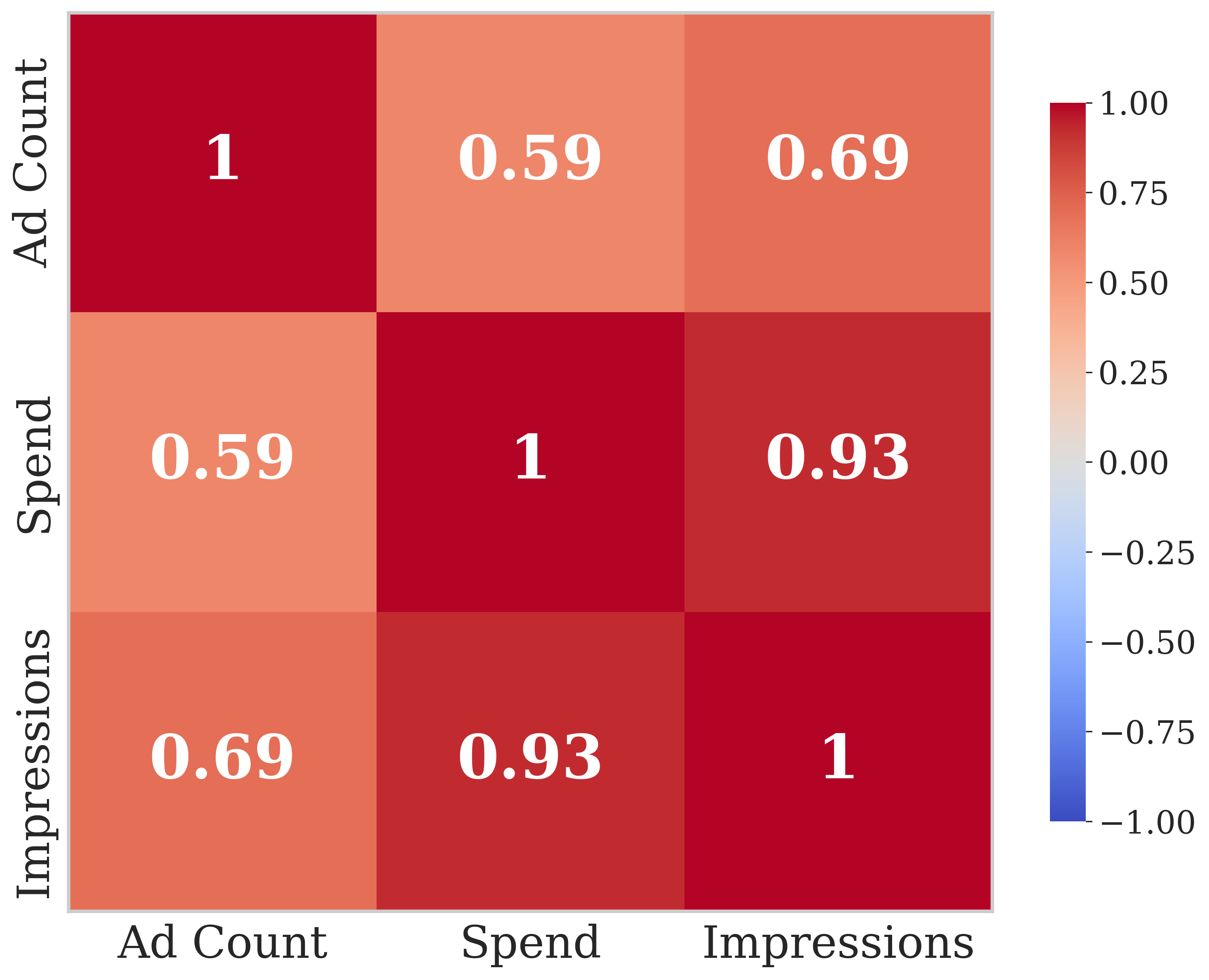}
        \caption{Spearman correlation matrix of stationary time series.}
        \label{fig:spearman_corr}
    \end{subfigure}
    \caption{Correlation matrices of stationary (differenced) time series. (a) Pearson correlation matrix showing strong positive relationships between all metrics, with the strongest correlation (0.94) between spend and impressions. (b) Spearman correlation matrix exhibiting similar patterns but with some differences in magnitude, suggesting potential non-linear relationships particularly involving ad count.}
    \label{fig:correlation_matrices}
\end{figure}

Figure \ref{fig:pearson_corr} visualizes the Pearson correlation coefficients between our differenced (stationary) time series. The correlation matrix reveals strong positive relationships between all three metrics, with the strongest correlation (r = 0.94) observed between daily changes in spend and impressions.

Figure \ref{fig:spearman_corr} presents the Spearman rank correlation coefficients for the same data. This non-parametric measure assesses monotonic relationships without assuming linearity. The pattern of correlations is similar to the Pearson coefficients, though with some notable differences. The Spend-Impressions correlation remains extremely strong ($\rho = 0.93$), essentially identical to the Pearson coefficient, indicating a highly consistent relationship. However, correlations involving Ad Count show more pronounced differences between Pearson and Spearman metrics (particularly Spend-Ad Count, dropping from $r = 0.76$ to $\rho = 0.59$), suggesting potential non-linear components in these relationships.

\subsection{Multiple Hypothesis Testing Correction}
\label{a:bonferroni}
When conducting multiple statistical tests, the probability of obtaining false positive results increases with each additional test. To control for this family-wise error rate, we applied the Bonferroni correction method, which adjusts the significance threshold by dividing the standard alpha level (0.05) by the number of tests performed.

Our analysis included a total of 12 correlation tests across the paper:

\begin{enumerate}
    \item \textbf{Descriptive Statistics ($\S$\ref{s:descriptive})}
    \begin{itemize}
        \item Pearson and Spearman correlations between ad expenses and impression numbers
        \item Pearson and Spearman correlations between ad duration and expenses
        \item Pearson and Spearman correlations between ad duration and impressions
    \end{itemize}

    \item \textbf{Temporal Analysis ($\S$\ref{ss:temporal})}
    \begin{itemize}
        \item Pearson and Spearman correlations between differenced spend and impressions
        \item Pearson and Spearman correlations between differenced ad count and impressions
        \item Pearson and Spearman correlations between differenced spend and ad count
    \end{itemize}
\end{enumerate}

The Bonferroni-corrected significance threshold was $\frac{0.05}{12} = 0.0042$. All our reported correlations have p-values substantially lower than this adjusted threshold, with most being below $10^{-9}$, indicating that the observed relationships remain statistically significant even after controlling for multiple comparisons.

Additionally, our analysis included diagnostic tests for time series properties:
\begin{enumerate}
    \item KPSS tests for stationarity (6 tests total: 3 on original series, 3 on differenced series)
    \item Mann-Kendall test for trend (1 test)
\end{enumerate}

These diagnostic tests were conducted sequentially as prerequisite checks for our correlation analyses rather than simultaneous hypothesis tests. The Mann-Kendall test ($p = 8.86 \times 10^{-35}$) remains highly significant under any reasonable correction. The KPSS test results for our original series (all $p = 0.01$) and differenced series (all $p = 0.10$) represent boundary values from the statistical implementation rather than exact p-values, and their interpretation remains consistent regardless of correction.

\section{Mining-Related Advertisement Analysis}
\label{a:mining}

\subsection{Methodology}

\new{To analyze mining-related political advertisements, we employed a keyword-based detection approach combined with regional and party-level aggregation. We identified mining-related content using a comprehensive set of keywords including: \textit{mine, mining, miner, resource, mineral, iron, ore, gas, oil, coal, drill, energy, fifo, export,} and \textit{dig}. All keywords were lemmatized and matched with ad texts.}

\new{Each advertisement was classified by its dominant region (the region receiving the highest percentage of targeting) and analyzed across three dimensions: regional focus, party focus, and party-region intersections. We calculated both advertisement count percentages and spending percentages to capture different aspects of mining-related campaign emphasis.}

\subsection{Key Findings}

\subsubsection{Party-Level Results}
\new{The analysis reveals stark differences in mining-related advertising focus across political parties. The Greens demonstrate the most intensive mining focus, allocating 50.03\% of their total advertising spend (\$304,702 out of \$609,006) to mining-related content—an order of magnitude higher than any other major party. The United Australia Party allocated 4.36\% of their spend (\$68,887 out of \$1,578,142) to mining topics, while the major parties showed much lower focus: Liberal (2.99\%) and Labor (1.59\%).}

\subsubsection{Regional Distribution}
\new{Mining-related advertising is disproportionately concentrated in smaller jurisdictions rather than major mining states. The Australian Capital Territory leads with 7.74\% of advertising spend on mining issues, followed by Tasmania (7.47\%) and South Australia (6.79\%). Notably, Queensland—despite its significant mining economy—ranks lowest among major states at 3.34\%, while Western Australia shows moderate focus at 6.51\%.}

\subsubsection{Strategic Targeting}
\new{The party-region analysis reveals distinct strategic approaches. The Greens maintain consistently high mining-related spending across all regions (ranging from 50-82\% of regional spend), suggesting a national anti-mining campaign strategy. The United Australia Party shows more targeted geographic focus, with highest mining-related spending in New South Wales (5.4\%) while maintaining minimal presence in other states.}

\subsubsection{Limitations of mining-related keyword analysis
}
\new{Our keyword-based approach provides a practical way to identify mining-related ads but may miss content using alternative phrasing, capture unrelated ads, ignore multimodal cues, and oversimplify regional targeting. Consequently, results should be seen as indicative rather than exhaustive.}

\end{document}